\documentclass[twocolumn]{revtex4}
\usepackage[dvips]{graphicx}
\usepackage{float, color,array}
\usepackage{amsfonts}

\begin{document}

\title{Spin current as a probe of the $\mathbb{Z}_2$-vortex topological transition in the classical Heisenberg antiferromagnet on the triangular lattice}

\author{K. Aoyama and H. Kawamura}

\date{\today}

\affiliation{Department of Earth and Space Science, Graduate School of Science, Osaka University, Osaka 560-0043, Japan
}

\begin{abstract}
We have theoretically investigated transport properties of the classical Heisenberg antiferromagnet on the triangular lattice in which a binding-unbinding topological transition of $\mathbb{Z}_2$ vortices is predicted to occur at a finite temperature $T_v$. It is shown by means of the hybrid Monte-Carlo and spin-dynamics simulations that the longitudinal spin-current conductivity exhibits a divergence at $T_v$, while the thermal conductivity only shows a monotonic temperature dependence with no clear anomaly at $T_v$. The significant enhancement of the spin-current conductivity is found to be due to the rapid growth of the spin-current-relaxation time toward $T_v$, which can be understood as a manifestation of the topological nature of the free $\mathbb{Z}_2$ vortex whose lifetime gets longer toward $T_v$. The result suggests that the spin-current measurement is a promising probe to detect the $\mathbb{Z}_2$-vortex topological transition which has remained elusive in experiments. 
\end{abstract}

\maketitle
In frustrated magnets, competitions between exchange interactions often result in a non-collinear magnetic state whose ordering wave vector may be commensurate or incommensurate with the underlying lattice structure. In the case of isotropic Heisenberg spins, such a non-collinear spin structure is invariant under any rotation in the three-dimensional spin space, so that the order parameter space has the topology of $SO(3)$. On the two-dimensional lattice, a point defect, namely, a vortex excitation, in the $SO(3)$ manifold is characterized by the topological number of $\mathbb{Z}_2$, and thus it is called a ``$\mathbb{Z}_2$ vortex''  \cite{Z2_Kawamura_84}. In contrast to an ordinary vortex having an integer topological number $\mathbb{Z}$, less is known about how the $\mathbb{Z}_2$ vortices affect magnetic properties of the system. In this letter, as a typical platform for the $\mathbb{Z}_2$ vortex, we consider the classical Heisenberg antiferromagnet on the triangular lattice, and investigate its transport properties, focusing on the role of the $\mathbb{Z}_2$-vortex excitations.     
 
\begin{figure}[t]
\includegraphics[scale=0.85]{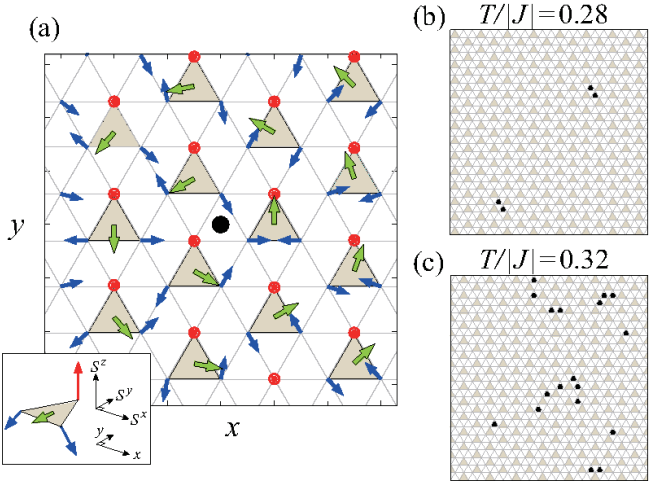}
\caption{(a) A schematically drawn $\mathbb{Z}_2$ vortex. The main panel shows spin and chirality vectors projected onto the two-dimensional lattice ($xy$) plane, where the central black dot represents a vortex core. The inset shows a zoomed three-dimensional view of each gray-colored triangle in the main panel, where a red (blue) arrow represents a spin vector ${\bf S}_i$ pointing upward (downward) without (with) in-plane components and a green one represents a chirality vector $\mbox{\boldmath $\kappa$}$. As the three spins at each gray-colored triangle constitute the 120$^\circ$ structure, $\mbox{\boldmath $\kappa$}$ does not have an out-of-plane component. Snapshots of the vortex-core distribution taken in the MC simulation below and above the $\mathbb{Z}_2$-vortex transition temperature $T_v/|J|=0.285$ are shown in (b) and (c), respectively, where black dots represent vortex cores. The definition of the $\mathbb{Z}_2$-vortex core is the same as that in Ref. \cite{Z2_Kawamura_84}. \label{fig:Z2}}
\end{figure}

The ground state of the triangular-lattice Heisenberg antiferromagnet with the nearest-neighbor (NN) exchange interaction $J$ \cite{Z2_Kawamura_84, Z2_Southern_95, Z2_Wintel_95, Z2_Kawamura_10, Z2_Kawamura_11} is the non-collinear 120$^\circ$ Neel state, in which three spins on each triangle, ${\bf S}_1$, ${\bf S}_2$, and ${\bf S}_3$, constitute additional degrees of freedom, a chirality vector $\mbox{\boldmath $\kappa$}=\frac{2}{3\sqrt{3}}({\bf S}_1\times{\bf S}_2+{\bf S}_2 \times {\bf S}_3 +{\bf S}_3\times{\bf S}_1)$. When the spin correlation develops over a lattice spacing at moderately high temperatures, the 120$^\circ$ spin structure is held in spatially local regions, e.g., elementary triangles. Such triangles having the local 120$^\circ$ structure and the associated chirality vector $\mbox{\boldmath $\kappa$}$ are building blocks of the $\mathbb{Z}_2$ vortex \cite{Z2_Kawamura_84}. A typical $\mathbb{Z}_2$ vortex is shown in Fig. \ref{fig:Z2}(a). In the three-component spin space, it forms a three-dimensionally oriented spin texture and can be viewed as a vortex formed by $\mbox{\boldmath $\kappa$}$. The topological object of the $\mathbb{Z}_2$ vortex is relevant to the phase transition in this system \cite{Z2_Kawamura_84, Z2_Southern_95, Z2_Wintel_95, Z2_Kawamura_10, Z2_Kawamura_11}.

As is well established, spins in the present system do not order except at $T=0$. In other words, the spin correlation length $\xi_s$ is finite at any finite temperature. In the middle '80s, Kawamura and Miyashita theoretically predicted that although spins are disordered with $\xi_s$ being finite, there exists a Kosterlitz-Thouless(KT)-type topological phase transition associated with binding-unbinding of the $\mathbb{Z}_2$ vortices \cite{Z2_Kawamura_84}. The $\mathbb{Z}_2$-vortex transition temperature $T_v$ is estimated to be $T_v/|J|\simeq 0.285$ via extensive Monte Carlo (MC) simulations \cite{Z2_Kawamura_10}. At lower temperatures $T<T_v$, all the $\mathbb{Z}_2$ vortices are paired up [see Fig. \ref{fig:Z2}(b)], while at higher temperatures $T>T_v$, dissociated free $\mathbb{Z}_2$ vortices can be found [see Fig. \ref{fig:Z2}(c)]. On approaching $T_v$ from above, the vortex density is reduced due to the vortex-pair annihilation, and correspondingly, the vortex correlation length $\xi_v$, which corresponds to the distance between {\it free} vortices, diverges toward $T_v$, whereas the spin correlation length $\xi_s$ remains finite \cite{Z2_Kawamura_10, Z2_Kawamura_11}. Once across $T_v$, the ergodicity is broken since the phase space is restricted only in the sector without free vortices. The low-temperature phase separated topologically from the ergodic disordered phase is sometimes called a ``spin gel'' state \cite{Z2_Kawamura_10, Z2_Kawamura_11, Sqomega_Okubo_jpsj_10}. 

In the triangular-lattice antiferromagnets NiGa$_2$S$_4$ \cite{NiGa2S4_Nakatsuji_science_05, NiGa2S4_Nambu_jpsj_06, NiGa2S4_Nambu_prl_08, NiGa2S4_Takeya_prb_08, NiGa2S4_MacLaughlin_prb_08, NiGa2S4_Yaouanc_prb_08, NiGa2S4_Yamaguchi_prb_08, NiGa2S4_Yamaguchi_jpsj_10, NiGa2S4_Nakatsuji_review_10, NiGa2S4_Stock_prl_10, NiGa2S4_Nambu_prl_15}, FeGa$_2$S$_4$ \cite{FeGa2S4_Zhao_prb_12, FeGa2S4_Reotier_prb_12}, NaCrO$_2$ \cite{NaCrO2_Olariu_pbl_06, NaCrO2_Hsieh_physicaB_08, NaCrO2_Hsieh_jpcs_08, NaCrO2_Hemmida_prb_09}, KCrO$_2$ \cite{KCrO2_Soubeyroux_79, KCrO2_Xiao_prb_13}, and AAg$_2$Cr[VO$_4$]$_2$ (A=K, Rb) \cite{AAg2CrV2O8_Tapp_prb_17}, a long-range magnetic order has not been observed down to the lowest temperature reachable in experiments, indicating the realization of the spin-gel state, and the possible existence of the $\mathbb{Z}_2$-vortex transition has extensively been discussed. Nevertheless, the $\mathbb{Z}_2$-vortex transition has remained elusive because static physical quantities such as the specific heat $C$ and the magnetic susceptibility $\chi_{m}$ exhibit only a weak essential singularity at $T_v$ \cite{Z2_Kawamura_10, Z2_Kawamura_11}. In this work, to propose a smoking-gun experiment to detect the transition, we examine dynamical physical quantities which may sensitively capture the dynamics characteristic of bound and unbound vortices. In this context, it was theoretically pointed out that the paired and free vortices show different characteristic features in the dynamical spin structure factor near $M$ and $K$ points of the Brillouin zone, respectively \cite{Sqomega_Okubo_jpsj_10}. Since in general, dynamical properties should also be reflected in transport phenomena, here, we theoretically investigate the conductivity of spin and thermal currents, putting particular emphasis on the spin transport which is nowadays becoming available as a probe to study magnetic fluctuations and excitations \cite{Spincurrent-mag_Frangou_16, Spincurrent-mag_Qiu_16, Spincurrent-mag_Wang_17, Spincurrent-mag_Frangou_17, Spincurrent-mag_Gladii_18, Spincurrent-mag_Ou_18, Spincurrent-mag_Li_19}.

In the low-temperature phase (spin-gel state) below $T_v$, spin and thermal currents should be carried by spin waves or magnons. At higher temperatures above $T_v$, thermally-activated free $\mathbb{Z}_2$ vortices may strongly affect the current relaxation, because the vortex as a topological object is generally robust against perturbations, resulting in a relatively long lifetime compared with the damping of the spin-wave mode.  
As we will demonstrate in this letter, this is actually the case for the spin-current relaxation which gets slower on cooling toward $T_v$, and as a result, the longitudinal spin-current conductivity grows up to diverge at $T_v$, serving as a distinct probe of the $\mathbb{Z}_2$-vortex transition.

The model Hamiltonian we consider is given by  
\begin{equation}\label{eq:Hamiltonian}
{\cal H} = -J \sum_{\langle i,j\rangle} {\bf S}_i \cdot {\bf S}_j ,
\end{equation}
and its dynamical properties is determined by the semi-classical equation of motion,
\begin{equation}\label{eq:Bloch}
\frac{d {\bf S}_i}{dt} = {\bf S}_i \times J\sum_{j\in N(i)}{\bf S}_j ,
\end{equation}
where ${\bf S}_i$ is a classical Heisenberg spin, $J<0$, $\langle i,j\rangle$ denotes the summation over all the NN pairs, and $N(i)$ denotes all the NN sites of $i$. Since Eq. (\ref{eq:Bloch}) is a classical analogue of the Heisenberg equation for the spin operator, all the static and dynamical magnetic properties intrinsic to the Hamiltonian (\ref{eq:Hamiltonian}) should be described by the combined use of Eqs. (\ref{eq:Hamiltonian}) and (\ref{eq:Bloch}). From the conservation of the magnetization and the energy, one can define the spin current ${\bf J}^\alpha_s$ and the thermal current ${\bf J}_{th}$ as follows \cite{SpinDyn_Huber_74, SpinDyn_Jencic_prb_15, MHall_Mook_prb_16, MHall_Mook_prb_17, Thermal_Huber_ptp_68, SpinDyn_Zotos_prb_05, SpinDyn_Kawasaki_67, SpinDyn_Sentef_07, SpinDyn_Pires_09, SpinDyn_Chen_13, TransportXXZ_AK_19},
\begin{equation}\label{eq:current_spin}
{\bf J}^\alpha_s = J\sum_{\langle i,j \rangle} \, \big({\bf r}_i-{\bf r}_j\big)({\bf S}_i \times {\bf S}_j)^\alpha,
\end{equation}
\begin{equation}\label{eq:current_th}
{\bf J}_{th} = \frac{J^2}{4}\sum_i  \sum_{j,k \in N(i)} \big( {\bf r}_j - {\bf r}_k \big) \, ({\bf S}_j\times{\bf S}_k) \cdot {\bf S}_i ,
\end{equation}
where $\alpha$ in ${\bf J}^\alpha_s$ denotes the spin component.  
One can see from Eqs. (\ref{eq:current_spin}) and (\ref{eq:current_th}) that ${\bf J}^\alpha_s$ and ${\bf J}_{th}$ are associated with the vector and scalar chiralities, respectively. Since the $\mathbb{Z}_2$ vortex is a texture formed by the vector chirality $\mbox{\boldmath $\kappa$}$, the spin transport is expected to be sensitive to the existence of the $\mathbb{Z}_2$ vortex.  

Within the linear response theory \cite{KuboFormular_Kubo_57}, one can define the spin-current conductivity $\sigma^s_{\mu \nu}$ and the thermal conductivity $\kappa_{\mu\nu}$ for the classical spin systems as follows \cite{SpinDyn_Kawasaki_67, SpinDyn_Zotos_prb_05, SpinDyn_Jencic_prb_15, MHall_Mook_prb_16, MHall_Mook_prb_17, TransportXXZ_AK_19},
\begin{eqnarray}\label{eq:conductivity_spin}
&& \sigma_{\mu \nu}^s = \frac{1}{T \, L^2} \int_0^\infty dt \, \big\langle J_{s,\nu}(0) \, J_{s,\mu}(t) \big\rangle, \\
&& \big\langle J_{s,\nu}(0) \, J_{s,\mu}(t) \big\rangle = \frac{1}{3} \sum_{\alpha=x,y,z}\big\langle J^\alpha_{s,\nu}(0) \, J^\alpha_{s,\mu}(t) \big\rangle, \nonumber
\end{eqnarray}
\begin{equation}\label{eq:conductivity_th}
\kappa_{\mu \nu} = \frac{1}{T^2 \, L^2} \int_0^\infty dt \, \big\langle J_{th,\nu}(0) \, J_{th,\mu}(t) \big\rangle,
\end{equation} 
where $L$ is a linear system size, $\langle O \rangle$ denotes the thermal average of a physical quantity $O$, and the spin-current conductivity $\sigma^s_{\mu \nu}$ is averaged over the three spin components because the spin space is isotropic in the present Heisenberg model. Noting that in Eq. (\ref{eq:Bloch}), time $t$ is measured in units of $|J|^{-1}$, it turns out that $\sigma^s_{\mu \nu}$ is a dimensionless quantity and $\kappa_{\mu \nu}$ has the dimension of $|J|$. 
As we take the lattice constant $a$ to be $a=1$, the total number of spin $N_{\rm spin}$ and $L$ is related by $N_{\rm spin}=L^2$.

In Eqs. (\ref{eq:current_spin}) and (\ref{eq:current_th}), the time evolutions of ${\bf J}^\alpha_s$ and ${\bf J}_{th}$ are determined microscopically by the spin-dynamics equation (\ref{eq:Bloch}). By using the second order symplectic method \cite{Symplectic_Krech_98, Sqomega_Okubo_jpsj_10, Symplectic_Furuya_11}, we numerically integrate Eq. (\ref{eq:Bloch}) typically up to $t=100\,|J|^{-1} \, - \, 800 \, |J|^{-1}$ with the time step $\delta t=0.01 \, |J|^{-1}$ and initial spin configurations generated by MC simulations. In this work, we performed 10-20 independent MC runs starting from different initial configurations under the periodic boundary conditions, and prepared 2000-4000 equilibrium spin configurations by picking up a spin snapshot every 1000 MC sweeps after 10$^5$ MC sweeps for thermalization, where one MC sweep consists of 1 heat-bath sweep and successive 10-30 over-relaxation sweeps. The thermal average is taken as the average over initial equilibrium spin configurations. We have checked that results are not altered if the 4th order Runge-Kutta method is used instead of the 2nd order symplectic method. By analyzing the system-size dependence of $\sigma^s_{\mu \nu}$ and $\kappa_{\mu\nu}$, we will discuss the temperature dependences of the conductivities in the thermodynamic limit ($L \rightarrow \infty$) of our interest.

\begin{figure}[t]
\includegraphics[scale=1.0]{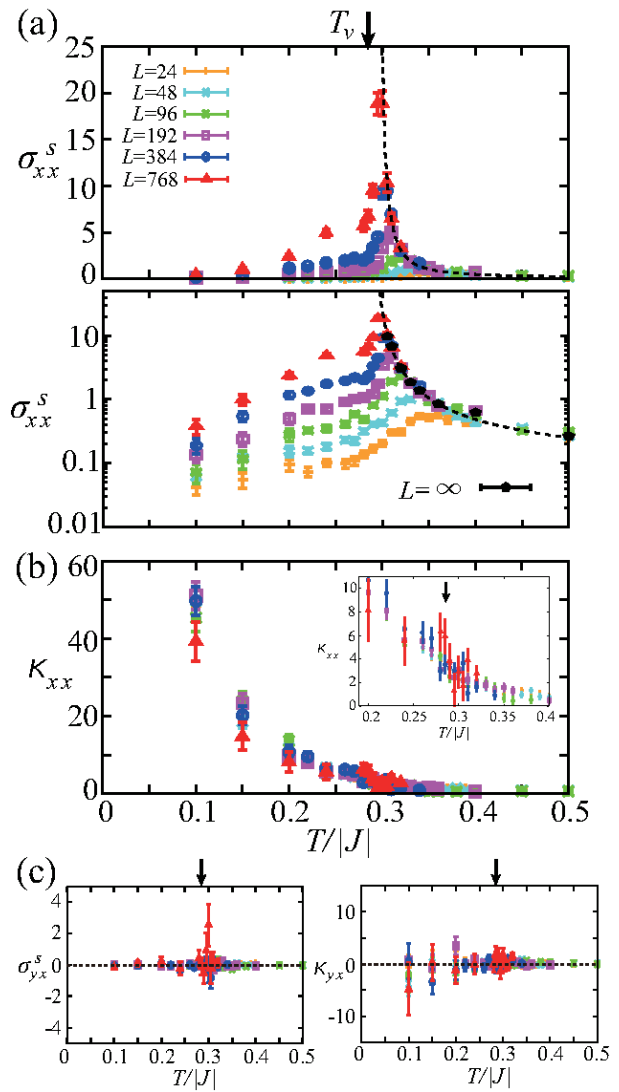}
\caption{The temperature dependence of the spin-current conductivity $\sigma^s_{\mu \nu}$ and the thermal conductivity $\kappa_{\mu \nu}$, where the longitudinal (transverse) components of $\sigma^s_{\mu \nu}$ and $\kappa_{\mu \nu}$ are shown in (a) [left panel of (c)] and (b) [right panel of (c)], respectively. $\kappa_{\mu \nu}$ is measured in units of $|J|$, whereas $\sigma^s_{\mu \nu}$ is a dimensionless quantity. A black arrow indicates the $\mathbb{Z}_2$-vortex transition temperature, $T_v/|J|\simeq 0.285$. In (a), the lower panel shows a semi-logarithmic plot of the upper panel together with black colored data representing $\sigma^s_{xx}$ values in the thermodynamic limit of $L \rightarrow \infty$, and a dashed curve represents the $\sigma^s_{xx}(T)$ curve obtained by fitting the $L \rightarrow \infty$ data (see the main text). In (b), the inset shows a zoomed view near $T_v$. \label{fig:trans}}
\end{figure}

Figure \ref{fig:trans} shows the longitudinal ($xx$) and transverse ($yx$) components of the spin-current conductivity $\sigma^s_{\mu \nu}$ and the thermal conductivity $\kappa_{\mu \nu}$ as a function of the temperature $T$, which are obtained for various system sizes ranging from $L=24$ to $L=768$. Since the result on the $yy$ ($xy$) component is qualitatively the same as that on the $xx$ ($yx$) one, only the latter is shown in Fig. \ref{fig:trans}, where the $x$ and $y$ directions are taken along the bond and off-bond directions of the triangular lattice, respectively [see Fig. \ref{fig:Z2} (a)]. As readily seen from Fig. \ref{fig:trans}, the longitudinal spin-current conductivity $\sigma_{xx}^s$ exhibits a divergent sharp peak near $T_v$, while the longitudinal thermal conductivity $\kappa_{xx}$ only shows a monotonic temperature dependence except a possible weak anomaly near $T_v$ which is, however, not evident at the precision of our numerical calculation. In both the spin and thermal transports, the transverse Hall response, $\sigma^s_{yx}$ and $\kappa_{yx}$, is absent at $2\sigma$ precision [see Fig. \ref{fig:trans} (c)]. 

We first discuss the low-temperature transport caused by magnons, which might be described by the linear-spin-wave theory (LSWT).
As shown in Fig. \ref{fig:trans} (b), $\kappa_{xx}$ increases monotonically toward $T=0$. As LSWT predicts $\kappa_{xx} \propto 1/\alpha_d$ with the magnon damping $\alpha_d$ \cite{Suppl}, the observed monotonic increase in $\kappa_{xx}$ can be understood as a result of the reduced scattering rate of magnons toward $T=0$, i.e., $\alpha_d \rightarrow 0$ \cite{TransportXXZ_AK_19}. Since the spin current should be carried by magnons as well, one may naively expect that $\sigma^s_{xx}$ increase toward $T=0$ similarly to $\kappa_{xx}$, but this does not seem to be the case for the numerically obtained $\sigma^s_{xx}$ [see Fig. \ref{fig:trans} (a)]. Also, in LSWT, the magnon-spin-current conductivity is calculated as $\sigma^s_{xx} \sim const \, T/\alpha_d + T \alpha_d \, \xi_s $ with $\xi_s \sim \exp\big[b_H |J|/T \big]$ \cite{Suppl}, suggesting that its temperature dependence is not so trivial because of the competition between $\alpha_d \rightarrow 0$ and $\xi_s \rightarrow \infty$. Such a situation is in sharp contrast to that of the unfrustrated Heisenberg antiferromagnet on the square lattice, in which $\sigma^s_{xx} \sim const \, T/\alpha_d + T \xi_s/\alpha_d $ is obtained, i.e., $\sigma^s_{xx}$ unambiguously increases in a monotonic manner toward $T=0$. This increasing behavior has been confirmed by numerical simulations \cite{TransportXXZ_AK_19}. 
In the present system, although the $T\rightarrow 0$ limit of $\sigma^s_{xx}$ remains unclarified, at least it is certain that the unusual low-temperature spin transport has its origin in the magnetic frustration.

Next, we discuss the significant enhancement of $\sigma^s_{xx}$ near $T_v$, which points to a strong association between the spin transport and the $\mathbb{Z}_2$-vortex transition.
As one can clearly see from the lower panel of Fig. \ref{fig:trans} (a), with increasing the system size $L$, the peak height in $\sigma_{xx}^s$ increases and the peak temperature approaches $T_v/|J|\simeq 0.285$ from above, suggesting that in the $L \rightarrow \infty$ limit, $\sigma^s_{xx}$ diverges at $T_v$. 
Since, at $T \gtrsim 0.3|J| > T_v$, $\sigma^s_{xx}$ saturates to a constant value as a function of the system size $L$ which corresponds to $\sigma^s_{xx}$ in the thermodynamic limit of $L \rightarrow \infty$. The $L \rightarrow \infty$ values of $\sigma^s_{xx}$ are represented by black symbols in the lower panel of Fig. \ref{fig:trans} (a). 

Now, we discuss the functional form characterizing the divergence of $\sigma^s_{xx}$ at $T_v$. 
Noting that the vortex correlation length $\xi_v$ grows up toward $T_v$ in the exponential form $\xi_v \sim \exp\big[A \big( \frac{|J|}{T-T_v} \big)^\alpha \big]$ with the estimated values of $\alpha=0.42$ and $A=0.84-0.97$ \cite{Z2_Kawamura_10}, we fit the $L \rightarrow \infty$ data of $\sigma^s_{xx}$ with the functional form of $b \exp\big[a\big( \frac{ |J|}{T-T_v} \big)^{0.42} \big]$. The resultant fitting function with the obtained values of $a=1.15 \pm 0.06$ and $b=0.028 \pm 0.007$ is represented by a dashed curve in Fig. \ref{fig:trans} (a). One can see that the obtained exponential form well characterizes the divergent behavior of $\sigma^s_{xx}$. 

To clarify the origin of the exponential divergence, we examine the temperature dependence of the time correlation function $\langle J_{s,x}(0) J_{s,x}(t) \rangle$ which involves the fundamental information about $\sigma^s_{xx}$ [see Eq. (\ref{eq:conductivity_spin})].  
\begin{figure}[t]
\includegraphics[scale=1.0]{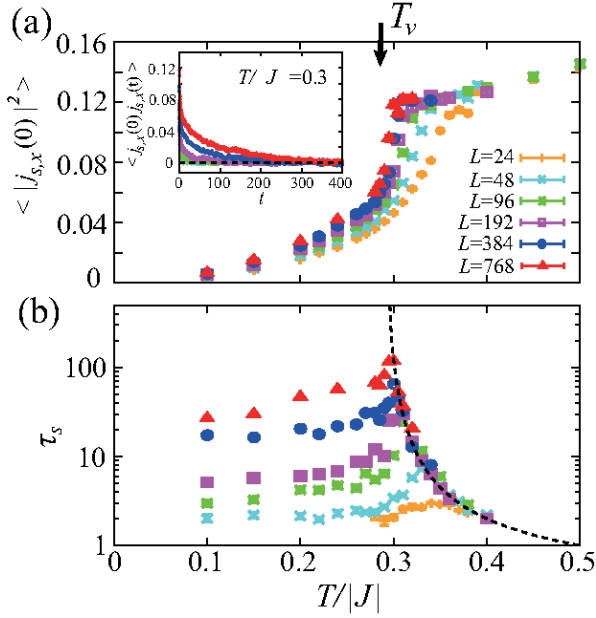}
\caption{The temperature dependence of the equal-time spin-current correlation $\langle | j_{s,x}(0)|^2 \rangle$ (a), and the spin-current relaxation time $\tau_s$ (b), which are measured in units of $|J|^2$ and $|J|^{-1}$, respectively. A black arrow indicates $T_v$ and a dashed curve in (b) represents an exponential function obtained by fitting the data above $T_v$ (see the main text). An inset of (a) shows the time correlation function of the spin current $\langle j_{s,x}(0) j_{s,x}(t) \rangle$ at $T/|J|=0.3$. \label{fig:detail}}
\end{figure}
The inset of Fig. \ref{fig:detail} (a) shows a typical example of the time correlation function normalized by the system size $\langle j_{s,x}(0) j_{s,x}(t) \rangle \equiv \langle J_{s,x}(0) J_{s,x}(t) \rangle/L^2$. As the time correlation decays exponentially in the form of $\exp[-t/\tau_s]$, one can define a characteristic time scale, namely, a spin-current-relaxation time $\tau_s$. Then, the time correlation function can roughly be written as $\langle j_{s,x}(0) j_{s,x}(t) \rangle \sim \langle |j_{s,x}(0)|^2 \rangle \, \exp[-t/\tau_s]$. By carrying out the integral over time in Eq. (\ref{eq:conductivity_spin}), one can estimate the longitudinal spin-current conductivity as $\sigma^s_{xx} \sim T^{-1} \, \tau_s \, \langle | j_{s,x}(0)|^2 \rangle$. Figure \ref{fig:detail} shows the temperature dependences of $\langle |j_{s,x}(0)|^2 \rangle$ and $\tau_s$, where $\tau_s$ is extracted by fitting the long-time tail of $\langle j_{s,x}(0) j_{s,x}(t) \rangle$ with $\exp[-t/\tau_s]$. 
One can see that on approaching $T_v$ from above, $\tau_s$ is significantly enhanced, while $\langle | j_{s,x}(0)|^2 \rangle$ exhibits a weaker anomaly. The functional type characterizing the steep increase in $\tau_s$ is also an exponential one. By fitting the data at $T/|J| \gtrsim 0.3$ with $ \tilde{b} \exp\big[\tilde{a}  \big( \frac{|J|}{T-T_v} \big)^{0.42} \big]$, we obtain $\tilde{a}=1.21 \pm 0.05$ and $\tilde{b}=0.10 \pm 0.02$. The extrapolated $\tau_s(T)$ curve represented by a dashed curve in Fig. \ref{fig:detail} (b) well characterizes the numerically obtained divergent behavior of $\tau_s$. As $\sigma^s_{xx}$ is related to $\tau_s$ and $\langle | j_{s,x}(0)|^2 \rangle$ via $\sigma^s_{xx}\sim T^{-1} \, \tau_s \, \langle | j_{s,x}(0)|^2 \rangle$, it turns out that the divergent behavior in $\sigma^s_{xx}$ originates from the exponential rapid growth of $\tau_s$ toward $T_v$. Actually, the obtained values of $a$ and $\tilde{a}$ are close to each other.

Here, we provide the physical interpretation of the above result. On cooling toward $T_v$, the inter-free-vortex distance $\xi_v$ increases, so that a free $\mathbb{Z}_2$ vortex wanders for a longer time until it collides with an other free $\mathbb{Z}_2$ vortex to be pair-annihilated. 
Since the vortex motion is not ballistic but rather diffusive \cite{diffusive_suppl}, the vortex lifetime $\tau_{v}$ could be estimated roughly as $\tau_{v} \propto \xi_v^2 \sim \exp\big[ 2A \big( \frac{|J|}{T-T_v} \big)^\alpha \big]$, so that $\tau_{v}$ should get longer in the exponential form toward $T_v$ with $ 2A \simeq 1.68-1.94$ which is comparable to $a$ and $\tilde{a}$.     
Since the two time-scales, $\tau_s$ and $\tau_{v}$, develop toward $T_v$ in almost the same manner as a function of the temperature, and furthermore, $\sigma^s_{xx}$ is proportional to $\tau_s$, we could conclude that the divergent peak at $T_v$ in the $\sigma^s_{xx}$ curve is attributed to the topological excitations of the long lifetime free $\mathbb{Z}_2$ vortices.

Finally, we address experimental aspects to detect the divergent enhancement of $\sigma^s_{\mu\mu}$ at $T_v$. Since a single crystal is necessary for transport experiments, a good candidate material in this respect would be NiGa$_2$S$_4$ \cite{NiGa2S4_Nakatsuji_science_05, NiGa2S4_Nambu_jpsj_06, NiGa2S4_Nambu_prl_08, NiGa2S4_Takeya_prb_08, NiGa2S4_MacLaughlin_prb_08, NiGa2S4_Yaouanc_prb_08, NiGa2S4_Yamaguchi_prb_08, NiGa2S4_Yamaguchi_jpsj_10, NiGa2S4_Nakatsuji_review_10, NiGa2S4_Stock_prl_10, NiGa2S4_Nambu_prl_15}. In the antiferromagnetic insulator NiGa$_2$S$_4$, although spins do not order down to the lowest temperature, a weak but clear transition-like anomaly, which may be attributed to the binding-unbinding of the $\mathbb{Z}_2$ vortices, has been observed at $T^\ast$ slightly below the specific-heat broad peak temperature. When the non-local measurement of the spin current \cite{nonlocal_Kajiwara_10, nonlocal_Giles_15, nonlocal_Cornelissen_15} is done on NiGa$_2$S$_4$, it is expected that a significant enhancement of $\sigma^s_{\mu\mu}$, i.e., a gigantic signal in the inverse spin Hall detector or a very-long-distance transport of spin information, be observed at $T^\ast$ as a distinct evidence for the $\mathbb{Z}_2$-vortex transition \cite{interface_suppl}. 
We emphasize that only $\sigma^s_{\mu\mu}$ diverges at $T_v$, while the static quantities $C$ and $\chi_m$ do not, which is in contrast to a ferromagnet where $C$ and $\chi_m$ as well as $\sigma^s_{\mu\mu}$ exhibit critical behaviors at the transition \cite{SpinDyn_Bennett_65, SpinDyn_Kawasaki_67, SpinDyn_Kawasaki_68, SpinDyn_Hohenberg_77}.
Such a characteristic spin-transport phenomenon may be observed also in FeGa$_2$S$_4$ \cite{FeGa2S4_Zhao_prb_12, FeGa2S4_Reotier_prb_12}, NaCrO$_2$ \cite{NaCrO2_Olariu_pbl_06, NaCrO2_Hsieh_physicaB_08, NaCrO2_Hsieh_jpcs_08, NaCrO2_Hemmida_prb_09}, and KCrO$_2$ \cite{KCrO2_Soubeyroux_79, KCrO2_Xiao_prb_13} in which a putative $\mathbb{Z}_2$-vortex anomaly has been reported, so that further experimental study including single-crystal growth on these compounds is strongly awaited.
We note that in the present system with finite $\xi_s$, perturbative interactions including magnetic anisotropy $D$ and the interlayer coupling $J'_{3D}$ which may exist in real materials are irrelevant as long as their effective energy scale, e.g., $D (J'_{3D}) \xi_s^2$, is smaller than $k_{\rm B} T_v$, while in the KT transition, $\xi_s$ diverges and thereby, they become inevitably relevant \cite{TransportXXZ_AK_19}.

In conclusion, we have theoretically shown that in the Heisenberg antiferromagnet on the triangular lattice, the longitudinal spin-current conductivity exhibits a divergence at the temperature of the $\mathbb{Z}_2$-vortex binding-unbinding topological transition. Such a significant enhancement of the spin transport is a smoking-gun experimental evidence for the so far elusive $\mathbb{Z}_2$-vortex transition which can potentially exist in a large variety of two-dimensional Heisenberg magnets possessing non-collinear spin correlations.

\begin{acknowledgments}
The authors thank K. Uematsu, S. Furuya, H. Adachi, and Y. Niimi for useful discussions. We are thankful to ISSP, the University of Tokyo and YITP, Kyoto University for providing us with CPU time. This work is supported by JSPS KAKENHI Grant Numbers JP16K17748, JP17H06137.
\end{acknowledgments}

\hspace{2cm}
\pagebreak

\onecolumngrid
\hspace{2cm}
\begin{center}
\textbf{\large Supplemental Material for ''Spin current as a probe of the $\mathbb{Z}_2$-vortex topological transition in the classical Heisenberg antiferromagnet on the triangular lattice''}\\[.2cm]
Kazushi Aoyama and Hikaru Kawamura \\[.2cm]
{\itshape Department of Earth and Space Science, Graduate School of Science, Osaka University, Osaka 560-0043, Japan}\\
(Dated: \today)\\[.5cm]
\end{center}

\section{Low-temperature properties of thermal and spin-current conductivities }
We will analytically investigate the temperature dependences of $\kappa_{\mu\nu}$ and $\sigma^s_{\mu \nu}$ based on the linear spin-wave theory (LSWT) which can be applied in the low-temperature limit. In Ref. \cite{TransportXXZ_AK_19}, we have taken the same analytical approach to derive $\kappa_{\mu\nu}$ and $\sigma^s_{\mu \nu}$ in the unfrustrated square-lattice antiferromagnet, and performed a calculation similar to that shown below. The difference between the present triangular-lattice and the previous square-lattice cases consists in the ordering wave vector ${\bf Q}$, the energy dispersion of the spin wave or the magnon $\varepsilon_{\bf q}$, and the magnon representation of the spin current, which results in the different temperature dependence of the spin-current conductivity $\sigma_{\mu\nu}^s$. Although the calculation method has already been addressed in Ref. \cite{TransportXXZ_AK_19}, here, we will provide all the steps to derive $\kappa_{\mu\nu}$ and $\sigma^s_{\mu \nu}$ for completeness. 

In LSWT, a low-temperature ordered state is a starting point. Although the present two-dimensional Heisenberg model does not exhibit a long-range magnetic order at any finite temperature, the spin-correlation length $\xi_s$ increases toward $T=0$ in the exponential form $\xi_s \sim a \exp[b_H |J|/T]$, where $a$ is a lattice constant and $b_H$ is a universal constant \cite{Heisenberg_Polyakov_75}. Thus, even at finite temperatures, the spin-wave expansions could be done locally within the regions smaller than $\xi_s$ \cite{MagnonDamping_Tyc_89}. We, therefore, introduce a lower cutoff in the momentum space which corresponds to the inverse spin-correlation length $\xi_s^{-1}$, and take the temperature dependence of $\xi_s$ into account. 

We will start from the theory of the corresponding quantum spin system, and then, take the classical limit of relevant physical quantities. By performing the spin-wave expansion, one can express the Hamiltoninan [Eq. (1) in the main text]
\begin{equation}\label{eq:Hamiltonian_suppl}
{\cal H} = -J \sum_{\langle i,j\rangle} {\bf S}_i \cdot {\bf S}_j ,
\end{equation}
and the spin and thermal currents [Eqs. (3) and (4) in the main text]
\begin{equation}\label{eq:j_sc}
{\bf J}^\alpha_s = J\sum_{\langle i,j \rangle} \, \big({\bf r}_i-{\bf r}_j\big)({\bf S}_i \times {\bf S}_j)^\alpha,
\end{equation}
\begin{equation}\label{eq:j_th}
{\bf J}_{th} = \frac{J^2}{4}\sum_i  \sum_{j,k \in N(i)} \big( {\bf r}_j - {\bf r}_k \big) \, ({\bf S}_j\times{\bf S}_k) \cdot {\bf S}_i ,
\end{equation}
in the magnon representation. Note that Eqs. (\ref{eq:j_sc}) and (\ref{eq:j_th}) for the classical spin systems can also be applied for quantum spin systems by merely replacing ${S}_i^\alpha$ with the associated spin operator $\hat{S}_i^\alpha$. We will calculate the conductivities $\kappa_{\mu \nu}$ and $\sigma^s_{\mu \nu}$ due to the magnon propagation, and discuss their temperature dependences in the classical limit.    
 
\subsection{Magnon representation}
Although our focus in the present paper is on the classical Heisenberg model, we start from the corresponding quantum spin system for convenience. By using the spin-wave expansions, we shall derive the magnon representation of the Hamiltonian (\ref{eq:Hamiltonian_suppl}) and the spin and thermal currents in Eqs. (\ref{eq:j_sc}) and (\ref{eq:j_th}).
Assuming that the 120$^\circ$ spin structure lies in the $xz$ plane in the spin space, we introduce the transformation from the laboratory frame to the rotated frame with $y$ being the rotation axis \cite{LSWT_rotation_Chubukov_jpcm_94, LSWT_rotation_Chernyshev_prb_09}, 
\begin{equation}
\left\{\begin{array}{l} 
S^x_i= \tilde{S}^z_i \sin(\theta_i) + \tilde{S}^x_i \cos(\theta_i), \\
S^z_i= \tilde{S}^z_i \cos(\theta_i) - \tilde{S}^x_i \sin(\theta_i), \\
S^y_i= \tilde{S}^y_i, 
\end{array} \right . \nonumber
\end{equation} 
where $\theta_i = {\bf Q}\cdot {\bf r}_i$ and ${\bf Q}=(\frac{4\pi}{3},0)$ is the ordering wave vector of the 120$^\circ$ structure. Then, the Hamiltonian reads as
\begin{equation}
{\cal H} = -\frac{J}{2} \sum_ i \sum_{j \in N(i)} \Big[ \tilde{S}^y_i \tilde{S}^y_j + \cos(\theta_i-\theta_j) \big( \tilde{S}^x_i \tilde{S}^x_j + \tilde{S}^z_i \tilde{S}^z_j \big) + \sin(\theta_i-\theta_j) \big( \tilde{S}^z_i \tilde{S}^x_j - \tilde{S}^x_i \tilde{S}^z_j \big)     \Big].
\end{equation} 
By using the Holstein-Primakoff transformation
\begin{equation}
\left\{\begin{array}{l} 
\tilde{S}^z_i = S- \hat{a}^\dagger_i \hat{a}_i , \\
\tilde{S}^x_i + i \tilde{S}^y_i = \sqrt{2S}\Big(1-\frac{\hat{a}^\dagger_i \hat{a}_i}{2S} \Big)^{\frac{1}{2}}\hat{a}_i = \sqrt{2S} \, \hat{a}_i +{\cal O}(S^{-\frac{1}{2}}) ,\\
\tilde{S}^x_i - i \tilde{S}^y_i = \sqrt{2S}\hat{a}^\dagger_i\Big(1-\frac{\hat{a}^\dagger_i \hat{a}_i}{2S} \Big)^{\frac{1}{2}} = \sqrt{2S} \, \hat{a}^\dagger_i+{\cal O}(S^{-\frac{1}{2}}) , \\
\end{array} \right . 
\end{equation}
with $\hat{a}^\dagger_i$ and $\hat{a}_i$ being respectively the bosonic creation and annihilation operators and the Fourier transformation of these operators 
\begin{equation}
\hat{a}^\dagger_i = \frac{1}{\sqrt{N}}\sum_{\bf q} \hat{a}^\dagger_{\bf q} e^{-i{\bf q}\cdot {\bf r}_i}, \quad \hat{a}_i = \frac{1}{\sqrt{N}}\sum_{\bf q} \hat{a}_{\bf q} e^{i{\bf q}\cdot {\bf r}_i},
\end{equation}
we obtain
\begin{eqnarray}
{\cal H} &=&\frac{1}{2}\sum_{\bf q}\Big[ A_{\bf q} \big( \hat{a}^\dagger_{\bf q}\hat{a}_{\bf q} + \hat{a}_{\bf q}\hat{a}^\dagger_{\bf q} \big)- B_{\bf q} \big(\hat{a}^\dagger_{\bf q}\hat{a}^\dagger_{-{\bf q}}+ \hat{a}_{\bf q}\hat{a}_{-{\bf q}} \big) \Big] + const. + {\cal O}(S^0), \nonumber\\
A_{\bf q} &=& -3JS \big( 1+\frac{1}{2} \gamma_{\bf q}\big), \qquad B_{\bf q} = -\frac{9}{2}JS \gamma_{\bf q}, \nonumber\\
\gamma_{\bf q} &=& \frac{1}{3}\Big[ \cos(q_x)+2\cos\big(\frac{1}{2}q_x \big)\cos\big( \frac{\sqrt{3}}{2}q_y \big) \Big].
\end{eqnarray}
By further using the Bogoliubov transformation
\begin{equation}
\left\{\begin{array}{l} 
\hat{a}_{\bf q} = u_{\bf q} \, \hat{b}_{\bf q} + v_{\bf q} \, \hat{b}^\dagger_{-{\bf q}}, \\
u_{\bf q}=u_{-{\bf q}} = \frac{1}{2}\Big[ \Big(\frac{A_{\bf q}+B_{\bf q}}{A_{\bf q}-B_{\bf q}} \Big)^{1/4} + \Big(\frac{A_{\bf q}-B_{\bf q}}{A_{\bf q}+B_{\bf q}} \Big)^{1/4}\Big], \nonumber\\
v_{\bf q}=v_{-{\bf q}} = \frac{1}{2}\Big[ \Big(\frac{A_{\bf q}+B_{\bf q}}{A_{\bf q}-B_{\bf q}} \Big)^{1/4} - \Big(\frac{A_{\bf q}-B_{\bf q}}{A_{\bf q}+B_{\bf q}} \Big)^{1/4}\Big],
\end{array} \right .  
\end{equation}
with $\hat{b}^\dagger_{\bf q}$ and $\hat{b}_{\bf q}$ being the creation and annihilation operators for magnons, one can diagonalize the above Hamiltonian for the $\hat{a}_{\bf q}$ magnons as follows: 
\begin{equation}\label{eq:Hamiltonian_mag}
{\cal H} \simeq \sum_{\bf q} \varepsilon_{\bf q} \, \hat{b}^\dagger_{\bf q}\hat{b}_{\bf q}, \qquad \varepsilon_{\bf q} = \sqrt{A_{\bf q}^2-B_{\bf q}^2}, 
\end{equation} 
where we have dropped constant and higher-order terms. As is well known, the magnon excitation energy $\varepsilon_{\bf q}$ becomes gapless at $\Gamma$ (${\bf q}=0$), $K$ (${\bf q}={\bf Q}$), and $K'$ (${\bf q}=-{\bf Q}$) points.

In the same manner, the thermal current in Eq. (\ref{eq:j_th}) can be expressed by the $\hat{b}_{\bf q}$ magnons as follows:
\begin{eqnarray}\label{eq:current_th_mag}
{\bf J}_{th} &=& \sum_{\bf q} \varepsilon_{\bf q} \, {\bf v}_{\bf q}  \, \hat{b}_{\bf q}^\dagger \hat{b}_{\bf q} + {\cal O}\big( S^{3/2} \big) , \nonumber\\
{\bf v}_{\bf q} &=& \nabla_{\bf q} \varepsilon_{\bf q}, 
\end{eqnarray}
whereas the spin current in Eq. (\ref{eq:j_sc}) as
\begin{eqnarray}\label{eq:current_spin_mag}
&&\left\{\begin{array}{l}
\displaystyle{ {\bf J}^x_s =  -\frac{JS}{4}i \sum_{\bf q} \big( \hat{\bf V}_{\bf q} - \hat{\bf V}^\dagger_{\bf q} \big) + {\cal O}\big( S^{1/2} \big)}, \\
\displaystyle{ {\bf J}^y_s =  -\frac{JS}{4} \sum_{\bf q} {\bf D}_{\bf q} (u_{\bf q}+v_{\bf q})^2 \big(  \hat{b}_{{\bf q}}^\dagger \hat{b}_{{\bf q}}+\hat{b}_{{\bf q}} \hat{b}_{{\bf q}}^\dagger + \hat{b}_{{\bf q}}^\dagger \hat{b}_{-{\bf q}}^\dagger + \hat{b}_{{\bf q}} \hat{b}_{-{\bf q}} \big) + {\cal O}\big( S^{1/2} \big) }, \\
\displaystyle{ {\bf J}^z_s =  \frac{JS}{4} \sum_{\bf q} \big( \hat{\bf V}_{\bf q} + \hat{\bf V}^\dagger_{\bf q} \big) + {\cal O}\big( S^{1/2} \big) }, \\
\end{array} \right . \nonumber\\
&& \hat{\bf V}_{\bf q} = {\bf C}_{\bf q}^1 \, \hat{b}_{{\bf q}+{\bf Q}}^\dagger \hat{b}_{-{\bf q}}^\dagger + {\bf C}_{\bf q}^2 \, \hat{b}_{{\bf q}} \hat{b}_{-{\bf q}-{\bf Q}} + {\bf C}_{\bf q}^3 \, \hat{b}_{{\bf q}} \hat{b}_{{\bf q}+{\bf Q}}^\dagger + {\bf C}_{\bf q}^4 \, \hat{b}_{-{\bf q}}^\dagger \hat{b}_{-{\bf q}-{\bf Q}} , \nonumber\\
&& {\bf C}^1_{\bf q} = ( {\bf C}_{\bf q}-{\bf C}_{{\bf q}+{\bf Q}}) (u_{{\bf q}+{\bf Q}}u_{{\bf q}}-v_{{\bf q}+{\bf Q}}v_{{\bf q}}) - 2({\bf C}_{{\bf q}}+{\bf C}_{{\bf q}+{\bf Q}}) u_{{\bf q}+{\bf Q}}v_{{\bf q}}, \nonumber\\
&& {\bf C}^2_{\bf q} = -( {\bf C}_{\bf q}-{\bf C}_{{\bf q}+{\bf Q}}) (u_{{\bf q}+{\bf Q}}u_{{\bf q}}-v_{{\bf q}+{\bf Q}}v_{{\bf q}}) - 2({\bf C}_{{\bf q}}+{\bf C}_{{\bf q}+{\bf Q}}) v_{{\bf q}+{\bf Q}}u_{{\bf q}}, \nonumber\\
&& {\bf C}^3_{\bf q} = ( {\bf C}_{\bf q}-{\bf C}_{{\bf q}+{\bf Q}}) (u_{{\bf q}+{\bf Q}}v_{{\bf q}}-v_{{\bf q}+{\bf Q}}u_{{\bf q}}) - 2({\bf C}_{{\bf q}}+{\bf C}_{{\bf q}+{\bf Q}}) u_{{\bf q}+{\bf Q}}u_{{\bf q}}, \nonumber\\
&& {\bf C}^4_{\bf q} = -( {\bf C}_{\bf q}-{\bf C}_{{\bf q}+{\bf Q}}) (u_{{\bf q}+{\bf Q}}v_{{\bf q}}-v_{{\bf q}+{\bf Q}}u_{{\bf q}}) - 2({\bf C}_{{\bf q}}+{\bf C}_{{\bf q}+{\bf Q}}) v_{{\bf q}+{\bf Q}}v_{{\bf q}}, \nonumber\\
&& {\bf C}_{\bf q} = \Big( \sin(q_x)+\sin\big( \frac{1}{2}q_x\big)\cos\big( \frac{\sqrt{3}}{2}q_y\big), \, \sqrt{3}\cos\big( \frac{1}{2}q_x\big)\sin\big( \frac{\sqrt{3}}{2}q_y\big) \Big), \nonumber\\
&& {\bf D}_{\bf q} = \sqrt{3}\Big(-\cos(q_x)+\cos\big( \frac{1}{2}q_x\big)\cos\big( \frac{\sqrt{3}}{2}q_y\big), \, - \sqrt{3} \sin\big( \frac{1}{2}q_x\big)\sin\big( \frac{\sqrt{3}}{2}q_y\big) \Big).
\end{eqnarray}

\subsection{Thermal and spin-current conductivities}  
Now, we shall calculate the thermal and spin-current conductivities which in the classical spin systems, are obtained from the time-correlation of the associated currents.
In order to calculate the thermal average of the time correlation, it is convenient to start from the quantum mechanical system and take the classical limit afterwards. In the quantum mechanical system, by introducing a response function
\begin{equation}
Q^a_{\mu\nu}(i\omega_n) = -\frac{1}{L^2}\int_0^{1/T}\big\langle T_{\tau} J_{a,\mu}(\tau)J_{a,\nu}(0) \big\rangle \, e^{i \omega_n \, \tau} d\tau \nonumber
\end{equation} 
with the bosonic Matsubara frequency $\omega_n=2\pi n T$, $\kappa_{\mu\nu}$ and $\sigma^s_{\mu\nu}$ are given by
\begin{eqnarray}\label{eq:conductivity_quantum}
\kappa_{\mu\nu} &=&\frac{1}{T} i \frac{d \, Q^{th,R}_{\mu\nu}(\omega)}{d \, \omega} \Big|_{\omega =0}, \nonumber\\
\sigma^s_{\mu\nu} &=& i \frac{d \, Q^{s,R}_{\mu\nu}(\omega)}{d \, \omega} \Big|_{\omega =0},
\end{eqnarray}
where $Q^{a,R}_{\mu\nu}(\omega) = Q^a_{\mu\nu}(\omega + i 0) $.

We first calculate the thermal conductivity $\kappa_{\mu\nu}$. Since the magnon representation of thermal current in Eq. (\ref{eq:current_th_mag}) takes the same form as the one for the unfrustrated square-lattice antiferromagnet, the calculation of $\kappa_{\mu\nu}$ performed below is the same as that for the unfrustrated system in Ref.\cite{TransportXXZ_AK_19}, except the concrete form of the summation over ${\bf q}$ in Eq. (\ref{eq:qsum}). For the thermal current carried by the magnons in Eq. (\ref{eq:current_th_mag}), the response function $Q^{th}_{\mu\nu}(i\omega_n)$ is given by \cite{book_AGD}
\begin{eqnarray}
Q^{th}_{\mu\nu}(i\omega_n) &=& \frac{-1}{L^2}\sum_{\bf q} \varepsilon_{\bf q}^2 v_{{\bf q},\mu} \, v_{{\bf q},\nu} \, T\sum_{\omega_m}{\cal D}_{\bf q}(i\omega_m){\cal D}_{\bf q}(i\omega_m+i\omega_n) \nonumber\\
&=& \frac{-1}{L^2}\sum_{\bf q} \varepsilon_{\bf q}^2 v_{{\bf q},\mu} \,  v_{{\bf q},\nu} \int_{-\infty}^{\infty}\frac{dx}{2\pi i} \big[ {\cal D}^R_{\bf q}(x)-{\cal D}^A_{\bf q}(x) \big] \big[ {\cal D}^R_{\bf q}(x+i\omega_n) + {\cal D}^A_{\bf q}(x-i\omega_n) \big] \, f_{\rm B}(x) ,
\end{eqnarray}
where $f_{\rm B}(x)=(e^{x/T}-1)^{-1}$ is the Bose-Einstein distribution function, and ${\cal D}^R_{\bf q}(x)$ (${\cal D}^A_{\bf q}(x)=\big[{\cal D}^R_{\bf q}(x)\big]^*$) is the retarded (advanced) magnon Green's function obtained by analytic continuation $i\omega_m \rightarrow \omega + i0$ in the temperature Green's function ${\cal D}_{\bf q}(i\omega_m)$ defined by 
\begin{equation}
{\cal D}_{\bf q}(\tau) = -\big\langle T_{\tau} \hat{b}_{\bf q}(\tau)\hat{b}_{\bf q}^\dagger(0) \big\rangle = T\sum_{\omega_m} {\cal D}_{\bf q}(i\omega_m) \, e^{-i\omega_m \tau}.
\end{equation} 
With use of Eq. (\ref{eq:conductivity_quantum}), the thermal conductivity in the quantum system is formally expressed as
\begin{equation}\label{eq:conductivity_quantum_th}
\kappa_{\mu\nu}=\frac{T^{-1}}{4\pi L^2}\int_{-\infty}^\infty dx \sum_{\bf q} \varepsilon_{\bf q}^2 \, v_{{\bf q},\mu} \, v_{{\bf q},\nu} f_{\rm B}'(x)\big[ {\cal D}^R_{\bf q}(x)-{\cal D}^A_{\bf q}(x) \big]^2.
\end{equation}
Here, the magnon Green's function ${\cal D}_{\bf q}^R(x)$ is given by
\begin{equation}\label{eq:Green_mag}
{\cal D}_{\bf q}^R(x) = \frac{1}{x-\varepsilon_{\bf q}+i \alpha_d \, x} = \big[ {\cal D}_{\bf q}^A(x) \big]^\ast,
\end{equation}
where the dimensionless coefficient $\alpha_d$ represents the magnon damping \cite{MagnonGreen_Yamaguchi_17, MagnonTrans_Tatara_15}. In general, the damping $\alpha_d$ originates from the interactions associated with spins in solids, so that it may be brought not only by the magnon-magnon scatterings but also, for example, by magnon-phonon scatterings. In the present work, however, the starting point is the spin Hamiltonian (\ref{eq:Hamiltonian_suppl}), and thus, $\alpha_d$ is of purely magnetic origin and brought by the magnon-magnon scatterings. 
Indeed, the magnon damping of the form $\alpha_d x$ proportional to the magnon energy $x$ in Eq. (\ref{eq:Green_mag}) has been derived elsewhere [see Eq. (4.11) in Ref. \cite{MagnonDamping_Tyc_89} and Eq. (4.19) in Ref. \cite{MagnonDamping_Harris_71}] for collinear ferromagnets and antiferromagnets by taking account only of the magnon-magnon scatterings. In the present triangular-lattice antiferromagnet possessing the non-collinear spin correlation, whether the magnon damping takes the same form or not is not clear, but here, we assume this form of the magnon damping for simplicity.

The thermal conductivity in the classical spin systems $\kappa_{\mu \nu}^{\rm cl}$ can be derived by substituting Eq. (\ref{eq:Green_mag}) into Eq. (\ref{eq:conductivity_quantum_th}) and taking the classical limit of 
\begin{equation}\label{eq:classical_limit}
f_{\rm B}(x) \rightarrow \frac{T}{x},
\end{equation}
or equivalently, $f_{\rm B}'(x)=-T/x^2$. Then, we obtain
\begin{equation}
\kappa_{\mu \nu}^{\rm cl} = \frac{1}{2L^2}\frac{1+\alpha_d^2}{\alpha_d}\sum_{\bf q}\frac{1}{\varepsilon_{\bf q}} \, v_{{\bf q},\mu} \, v_{{\bf q},\nu},
\end{equation}
where the equation
\begin{equation}\label{eq:integral}
\int_{-\infty}^\infty \frac{dx}{ \big[(x-\varepsilon_{\bf q})^2+(\alpha_d x)^2\big]^2 } = \frac{\pi}{2}\frac{1+\alpha_d^2}{\varepsilon_{\bf q}^3 \alpha_d^3}
\end{equation}
has been used. At this point, one may think that the temperature dependence of $\kappa_{\mu \nu}^{\rm cl}$ is dominated only by $\alpha_d$, but this is not the case for the present Heisenberg model. The additional temperature dependence due to the spin-correlation length $\xi_s$ comes in through the summation over ${\bf q}$. As we mentioned in the beginning of this supplemental material, $\xi_s$ enters in the form of the lower cutoff in the ${\bf q}$ space, i.e., $\xi_s^{-1} \leq |{\bf q}|$. Since the gapless excitations at ${\bf q}=0$ and $\pm {\bf Q}$ should contribute to the magnon transport, we replace the ${\bf q}$-summation with the following integral,
\begin{equation}\label{eq:qsum}
\sum_{\bf q} \, f_{\bf q} \simeq \sum_{{\bf q}^\ast = 0, \pm {\bf Q}} \frac{L^2}{(2\pi)^2} \int_0^{2\pi} d\phi_{\bf q} \int_{a/\xi_s}^{\Lambda} q \, dq \, f_{{\bf q}+{\bf q}^\ast}.
\end{equation}
As the original ${\bf q}$-summation is taken over the Brillouin zone, the upper cutoff $\Lambda$ for the integral with respect to $q=|{\bf q}|$ is of the order of unity, i.e., $\Lambda \sim {\cal O}(1)$.

As our focus is on the contributions coming from the gapless magnons, we will use the linear approximation which is valid near the gapless points,  
\begin{eqnarray}\label{eq:magenergy_app}
\varepsilon_{{\bf q}+{\bf q}^\ast} &\simeq& v_{{\bf q}^\ast} \, q, \nonumber\\
{\bf v}_{{\bf q}+{\bf q}^\ast}  &\simeq& v_{{\bf q}^\ast}\big(\cos\phi_{\bf q}, \sin\phi_{\bf q} \big), \nonumber\\
v_{{\bf q}^\ast} &=& \left\{ \begin{array}{l}
\frac{3\sqrt{3}}{2}|J|S \qquad ({\bf q}^\ast =0)\\
\frac{3\sqrt{3}}{2\sqrt{2}}|J|S \qquad ({\bf q}^\ast = \pm {\bf Q})\\
\end{array} \right. .
\end{eqnarray}
Note that the magnon velocities at $\Gamma$ (${\bf q}^\ast = 0$) and $K$, $K'$ (${\bf q}^\ast= \pm{\bf Q}$) points are different, i.e., $v_{0} > v_{\pm {\bf Q}}$. 

With the use of Eqs. (\ref{eq:qsum}) and (\ref{eq:magenergy_app}), we obtain
\begin{equation}\label{eq:conductivity_classical_th}
\kappa_{\mu \nu}^{\rm cl} \simeq \delta_{\mu,\nu}\frac{1}{8 \pi}\frac{1+\alpha_d^2}{\alpha_d} \big( v_{0} +2 \, v_{\bf Q} \big) \big( \Lambda - a/\xi_s \big) .
\end{equation}
Only the longitudinal components of the thermal conductivity $\kappa_{\mu \mu}^{\rm cl}$ are non-vanishing. Although the spin-correlation length $\xi_s$ rapidly increases toward $T=0$, such a temperature effect is irrelevant at lower temperatures because as one can see from Eq. (\ref{eq:conductivity_classical_th}), $\xi_s$ enters in $\kappa_{\mu\mu}^{\rm cl}$ in the form of $1/\xi_s$. When the magnon damping is sufficiently small such that $\alpha_d \ll 1$, it follows that $\kappa_{\mu\nu}^{\rm cl}\propto 1/\alpha_d$. In the $T \rightarrow 0$ limit, the magnon damping factor $\alpha_d$ should go to zero, so that correspondingly, $\kappa_{\mu\mu}^{\rm cl} \propto 1/\alpha_d$ increases toward $T=0$. It should be noted that in the integral over $q$, $\int_{a/\xi_s}^\Lambda dq \, q^x$, $\xi_s$ becomes relevant for a quantity satisfying $x \leq -1$. As we will see below, in contrast to the thermal conductivity with $x=0$, $\xi_s$ becomes relevant for the spin-current conductivity. 

Now, we shall calculate the spin-current conductivity $\sigma^s_{\mu\nu}$ based on Eq. (\ref{eq:conductivity_quantum}). In contrast to the thermal current, the magnon representation of the spin current in Eq. (\ref{eq:current_spin_mag}) takes the form different from the one for the unfrustrated square-lattice antiferromagnet \cite{TransportXXZ_AK_19}, so that, as we will see below, the associated conductivity $\sigma^s_{\mu\nu}$ in the present frustrated system exhibits a temperature dependence different from that in the unfrustrated system. As in the case of the thermal current, the response function of the spin current $Q^s_{\mu\nu}(i\omega_n)$ can be written as 
\begin{eqnarray}\label{eq:responsefnc}
Q^s_{\mu\nu}(i\omega_n) &=& Q^{s_{xz}}_{\mu\nu}(i\omega_n) + Q^{s_y}_{\mu\nu}(i\omega_n), \nonumber\\
Q^{s_{xz}}_{\mu\nu}(i\omega_n) &=& -\Big( \frac{JS}{4L} \Big)^2 \sum_{{\bf q}} \Big\{ \big[ {\bf C}_{\bf q}^1-{\bf C}_{\bf q}^2 \big]_\mu \, \big[ {\bf C}_{\bf q}^1-{\bf C}_{\bf q}^2 \big]_\nu \, F^-_{\bf q}(i \omega_n) + \big[ {\bf C}_{\bf q}^3-{\bf C}_{\bf q}^4 \big]_\mu \, \big[ {\bf C}_{\bf q}^3-{\bf C}_{\bf q}^4 \big]_\nu \, F^+_{\bf q}(i \omega_n) \Big\} , \nonumber\\
Q^{s_y}_{\mu\nu}(i\omega_n) &=& -\Big( \frac{JS}{4L} \Big)^2 \sum_{{\bf q}} 2\big[{\bf D}_{\bf q}\big]_\mu \big[{\bf D}_{\bf q}\big]_\nu (u_{\bf q}+v_{\bf q})^4  F^0_{\bf q}(i \omega_n) , \nonumber\\ 
F^\pm_{\bf q}(i \omega_n) &=& T\sum_{\omega_m} {\cal D}_{\bf q}(i\omega_m) \big[ {\cal D}_{{\bf Q}+ {\bf q}}(i\omega_n\pm i\omega_m) + {\cal D}_{{\bf Q}+{\bf q}}(-i\omega_n\pm i\omega_m) \big] \nonumber\\
&=& \int_{-\infty}^\infty \frac{dx}{2\pi i} f_{\rm B}(x)\Big\{ \big[ {\cal D}^R_{{\bf Q}+ {\bf q}}(\pm x+i \omega_n) + {\cal D}^A_{{\bf Q}+ {\bf q}}(\pm x-i \omega_n) \big] \big[ {\cal D}^R_{\bf q}(x)- {\cal D}^A_{\bf q}(x) \big]  \nonumber\\
&& \qquad\qquad\qquad \pm \big[ {\cal D}^R_{{\bf Q}+ {\bf q}}(\pm x) - {\cal D}^A_{{\bf Q}+ {\bf q}}(\pm x) \big] \big[ {\cal D}^R_{\bf q}(x+i\omega_n) + {\cal D}^A_{\bf q}(x-i\omega_n) \big]  \Big\}, \nonumber\\
F^0_{\bf q}(i \omega_n) &=& T\sum_{\omega_m} {\cal D}_{\bf q}(i\omega_m) \big[ {\cal D}_{{\bf q}}(i\omega_n+ i\omega_m) + {\cal D}_{{\bf q}}(-i\omega_n+ i\omega_m) + {\cal D}_{{\bf q}}(i\omega_n- i\omega_m) + {\cal D}_{{\bf q}}(-i\omega_n- i\omega_m) \big] \nonumber\\
&=& \int_{-\infty}^\infty \frac{dx}{2\pi i} f_{\rm B}(x)\Big\{ \big[ {\cal D}^R_{{\bf q}}( x+i \omega_n) + {\cal D}^A_{{\bf q}}(x-i \omega_n) + {\cal D}^R_{{\bf q}}(- x+i \omega_n) + {\cal D}^A_{{\bf q}}(- x-i \omega_n)\big] \big[ {\cal D}^R_{\bf q}(x)- {\cal D}^A_{\bf q}(x) \big] \nonumber\\
&&\qquad\qquad\qquad  + \big[ {\cal D}^R_{{\bf q}}(x) - {\cal D}^A_{{\bf q}}(x) - {\cal D}^R_{{\bf q}}(- x) + {\cal D}^A_{{\bf q}}(- x)\big] \big[ {\cal D}^R_{\bf q}(x+i\omega_n) + {\cal D}^A_{\bf q}(x-i\omega_n) \big] \Big\}. 
\end{eqnarray}
As in the present calculation, the 120$^\circ$ spin structure is assumed to lie in the $xz$ plane in the spin space, the spin current with its spin polarization in the $xz$ plane is not equivalent to the one with its polarization in the $y$ direction and the associated spin-current conductivities, $\sigma^{s_{xz}}_{\mu\nu}$ and $\sigma^{s_y}_{\mu\nu}$, are also the case. Thus, we will calculate $\sigma^{s_{xz}}_{\mu\nu}$ and $\sigma^{s_y}_{\mu\nu}$ separately. For this purpose, in Eq. (\ref{eq:responsefnc}), we decouple the response function into $Q^{s_{xz}}_{\mu\nu}(i\omega_n)$ and $Q^{s_y}_{\mu\nu}(i\omega_n)$ which correspond to the response functions of the spin current with its polarization in the $xz$ plane and the $y$ direction, respectively.
Then, the spin-current conductivity $\sigma^s_{\mu \nu}$ is formally written as
\begin{eqnarray}\label{eq:conductivity_quantum_spin}
&&\sigma^s_{\mu\nu} = \sigma^{s_{xz}}_{\mu\nu} + \sigma^{s_y}_{\mu\nu}, \nonumber\\
&&\sigma^{s_{xz}}_{\mu\nu} = \frac{-1}{2\pi}\Big( \frac{JS}{4L} \Big)^2\int_{-\infty}^\infty dx \sum_{{\bf q}} \, f_{\rm B}'(x) \big[ {\cal D}^R_{\bf q}(x)- {\cal D}^A_{\bf q}(x) \big]  \Big\{ \big[ {\bf C}_{\bf q}^1-{\bf C}_{\bf q}^2 \big]_\mu \, \big[ {\bf C}_{\bf q}^1-{\bf C}_{\bf q}^2 \big]_\nu  \big[ {\cal D}^R_{{\bf q}+{\bf Q}}(-x)- {\cal D}^A_{{\bf q}+{\bf Q}}(-x) \big] \\
&& \qquad \qquad -\big[ {\bf C}_{\bf q}^3-{\bf C}_{\bf q}^4 \big]_\mu \, \big[ {\bf C}_{\bf q}^3-{\bf C}_{\bf q}^4 \big]_\nu  \big[ {\cal D}^R_{{\bf q}+{\bf Q}}(x)- {\cal D}^A_{{\bf q}+{\bf Q}}(x) \big] \Big\} , \nonumber\\
&&\sigma^{s_y}_{\mu\nu} =  \frac{-2}{2\pi}\Big( \frac{JS}{4L} \Big)^2\int_{-\infty}^\infty dx \sum_{{\bf q}} \, f_{\rm B}'(x) \big[ {\cal D}^R_{\bf q}(x)- {\cal D}^A_{\bf q}(x) \big] \big[{\bf D}_{\bf q}\big]_\mu \big[{\bf D}_{\bf q}\big]_\nu (u_{\bf q}+v_{\bf q})^4 \big[  {\cal D}^R_{{\bf q}}(-x)- {\cal D}^A_{{\bf q}}(-x) - {\cal D}^R_{{\bf q}}(x)+ {\cal D}^A_{{\bf q}}(x)\big] . \nonumber
\end{eqnarray}
In the same manner as that for $\kappa_{\mu\nu}$, we take the classical limit of Eq. (\ref{eq:conductivity_quantum_spin}).
By substituting Eq. (\ref{eq:Green_mag}) into Eq. (\ref{eq:conductivity_quantum_spin}), taking the classical limit of $f_{\rm B}'(x)=-T/x^2$, and using Eq. (\ref{eq:integral}) and the formula
\begin{equation}
\int_{-\infty}^\infty \frac{dx}{ \big[(x-\varepsilon_{\bf q})^2+(\alpha_d x)^2\big]\big[(x+\varepsilon_{\bf q})^2+(\alpha_d x)^2\big] } = \frac{\pi}{2}\frac{1}{\varepsilon_{\bf q}^3 \alpha_d}, \nonumber
\end{equation}
we have the spin-current conductivity in the classical spin systems $\sigma^{s,{\rm cl}}_{\mu \nu}$ as follows:
\begin{eqnarray}\label{eq:conductivity_classical_spin_tmp}
\sigma^{s,{\rm cl}}_{\mu\nu} &=& \sigma^{s_{xz},{\rm cl}}_{\mu\nu} + \sigma^{s_y,{\rm cl}}_{\mu\nu}, \nonumber\\
\sigma^{s_{xz},{\rm cl}}_{\mu\nu} &=& \Big( \frac{JS}{4L} \Big)^2 2T\sum_{\bf q} \Big\{ \big[ {\bf C}_{\bf q}^1-{\bf C}_{\bf q}^2 \big]_\mu \, \big[ {\bf C}_{\bf q}^1-{\bf C}_{\bf q}^2 \big]_\nu \frac{\alpha_d}{\varepsilon_{\bf q}\varepsilon_{{\bf q}+{\bf Q}}(\varepsilon_{\bf q}+\varepsilon_{{\bf q}+{\bf Q}}) } \nonumber\\
&& + \big[ {\bf C}_{\bf q}^3-{\bf C}_{\bf q}^4 \big]_\mu \, \big[ {\bf C}_{\bf q}^3-{\bf C}_{\bf q}^4 \big]_\nu \frac{\alpha_d(1+\alpha_d^2)(\varepsilon_{\bf q}+\varepsilon_{{\bf q}+{\bf Q}})}{\varepsilon_{\bf q}\varepsilon_{{\bf q}+{\bf Q}}\big[(\varepsilon_{\bf q}-\varepsilon_{{\bf q}+{\bf Q}})^2+\alpha_d^2(\varepsilon_{\bf q}+\varepsilon_{{\bf q}+{\bf Q}})^2 \big] } \Big\}, \nonumber\\ 
\sigma^{s_y,{\rm cl}}_{\mu\nu} &=&\Big( \frac{JS}{4L} \Big)^2 2T\sum_{\bf q} \big[{\bf D}_{\bf q}\big]_\mu \big[{\bf D}_{\bf q}\big]_\nu (u_{\bf q}+v_{\bf q})^4 \frac{1+2\alpha_d^2}{\alpha_d}\frac{1}{\varepsilon_{\bf q}^3}. 
\end{eqnarray}
As our focus is on the small ${\bf q}$ region near the gapless points, we expand ${\bf q}$-dependent quantities appearing in Eq. (\ref{eq:conductivity_classical_spin_tmp}) with respect to ${\bf q}$, and obtain the leading order contributions as
\begin{eqnarray}\label{eq:spin_suppl}
{\bf C}_{{\bf q}+{\bf q}^\ast}^1-{\bf C}_{{\bf q}+{\bf q}^\ast}^2 &\simeq& 3\big(\frac{9}{2} \big)^{1/4}\big(\cos\phi_{\bf q}, \sin\phi_{\bf q} \big) \left\{ \begin{array}{l}
1 \qquad\qquad\quad ({\bf q}^\ast =0) \\
0 + {\cal O}(q^4) \quad ({\bf q}^\ast ={\bf Q}) \\
-1 \qquad\qquad ({\bf q}^\ast =-{\bf Q}) \\
\end{array} \right., \nonumber\\
{\bf C}_{{\bf q}+{\bf q}^\ast}^3-{\bf C}_{{\bf q}+{\bf q}^\ast}^4 &\simeq& 3\big(\frac{9}{2} \big)^{1/4}\big(\cos\phi_{\bf q}, \sin\phi_{\bf q} \big) \left\{ \begin{array}{l}
1 \qquad\qquad ({\bf q}^\ast = 0) \\
\big(\frac{2}{9} \big)^{1/4} \, q \quad ({\bf q}^\ast ={\bf Q}) \\
1 \qquad\qquad ({\bf q}^\ast =-{\bf Q}) \\
\end{array} \right., \nonumber\\
{\bf D}_{\bf q} &\simeq& q^2 \, \big( \frac{3\sqrt{3}}{8}\big[\cos^2\phi_{\bf q}-\sin^2\phi_{\bf q} \big] , \,  -\frac{3}{4}\cos \phi_{\bf q} \, \sin\phi_{\bf q}  \big), \nonumber\\
{\bf D}_{{\bf q}\pm {\bf Q}} &\simeq& \pm q \, \big( -\frac{3}{2}\cos\phi_{\bf q}, \, \frac{3\sqrt{3}}{4}\sin\phi_{\bf q} \big), \nonumber\\ 
(u_{{\bf q}+{\bf q}^\ast}+v_{{\bf q}+{\bf q}^\ast})^4 &=& \frac{A_{{\bf q}+{\bf q}^\ast}+B_{{\bf q}+{\bf q}^\ast}}{A_{{\bf q}+{\bf q}^\ast}-B_{{\bf q}+{\bf q}^\ast}}\simeq \left\{ \begin{array}{l}
12/q^2 \qquad ({\bf q}^\ast =0) \\
q^2/6 \qquad ({\bf q}^\ast =\pm{\bf Q}) \\
\end{array} \right. .
\end{eqnarray}
By substituting Eq. (\ref{eq:spin_suppl}) into Eq. (\ref{eq:conductivity_classical_spin_tmp}) and performing the ${\bf q}$-summation with the use of Eq. (\ref{eq:qsum}), we obtain
\begin{eqnarray}\label{eq:conductivity_classical_spin}
\sigma^{s_{xz},{\rm cl}}_{\mu\nu} &\simeq& \delta_{\mu,\nu}\frac{J^2S^2}{16 \pi} T \, \bigg[ \big(\frac{\xi_s}{a}-\frac{1}{\Lambda} \big) \frac{27}{\sqrt{2}}\Big( \frac{\alpha_d}{v_0 v_{\bf Q}(v_0+v_{\bf Q} )} + \frac{\alpha_d(1+\alpha_d^2)(v_0+v_{\bf Q})}{v_0v_{\bf Q}\big[ (v_0-v_{\bf Q} )^2+\alpha_d^2(v_0+v_{\bf Q} )^2\big] }\Big) + \big( \Lambda - \frac{a}{\xi_s} \big) \frac{9}{4} \frac{1+\alpha_d^2}{\alpha_d}\frac{1}{v_{\bf Q}^3}  \bigg], \nonumber\\
\sigma^{s_y,{\rm cl}}_{\mu\nu} &\simeq& \delta_{\mu,\nu}\frac{J^2S^2}{16 \pi} T \, \bigg[ \big( \Lambda - \frac{a}{\xi_s} \big)\big( 3\delta_{\mu,x}+\delta_{\mu,y}\big)\frac{27}{32}\frac{1+2\alpha_d^2}{\alpha_d}\frac{1}{v_0^3} + \Big\{ \Lambda^3 - \big( \frac{a}{\xi_s}\big)^3 \Big\} \big( 4\delta_{\mu,x}+3\delta_{\mu,y}\big)\frac{1}{16}\frac{1+2\alpha_d^2}{\alpha_d}\frac{1}{v_{\bf Q}^3} \bigg] .
\end{eqnarray}
At low temperatures such that the magnon damping is sufficiently small, i.e., $\alpha \ll 1$, Eq. (\ref{eq:conductivity_classical_spin}) is reduced to
\begin{eqnarray}\label{eq:conductivity_classical_spin_final}
\sigma^{s_{xz},{\rm cl}}_{\mu\nu} &\simeq& \delta_{\mu,\nu}\frac{J^2S^2}{16 \pi} \Big[ \, \frac{T}{\alpha_d} \, c_{\sigma 1}  + \alpha_d \, T  \big(\frac{\xi_s}{a}-\frac{1}{\Lambda} \big) \, c_{\sigma 2}\Big], \nonumber\\
c_{\sigma 1} &=& \frac{1}{v_{\bf Q}^3} \frac{9}{4}\big( \Lambda - \frac{a}{\xi_s} \big), \nonumber\\
c_{\sigma 2} &=&  \frac{27}{\sqrt{2}}\Big( \frac{1}{v_0 v_{\bf Q}(v_0+v_{\bf Q} )} + \frac{v_0+v_{\bf Q}}{v_0v_{\bf Q}(v_0-v_{\bf Q} )^2 }\Big), \nonumber\\
\sigma^{s_y,{\rm cl}}_{\mu\nu} &\simeq& \delta_{\mu,\nu}\frac{J^2S^2}{16 \pi} \frac{T}{\alpha_d} \, c'_{\sigma 1}, \nonumber\\
c'_{\sigma 1} &=& \frac{1}{v_{\bf Q}^3} \Big[ \frac{1}{16} \Big\{ \Lambda^3 - \big( \frac{a}{\xi_s}\big)^3 \Big\} \big( 4\delta_{\mu,x}+3\delta_{\mu,y}\big)\Big] + \frac{1}{v_0^3} \frac{27}{32}\big( \Lambda - \frac{a}{\xi_s} \big)\big( 3\delta_{\mu,x}+\delta_{\mu,y}\big), \nonumber\\
\end{eqnarray}
The transverse component is absent and the longitudinal spin-current conductivity exhibits a temperature dependence of the form $\sigma^{s_{xz},{\rm cl}}_{\mu\mu} \sim const \,  T/\alpha_d + T \,\alpha_d \, \xi_s$ and $\sigma^{s_y,{\rm cl}}_{\mu\mu} \sim const \,  T/\alpha_d $. $\sigma^{s_{xz},{\rm cl}}_{\mu\mu}$ involves the spin correlation length $\xi_s$, while $\sigma^{s_y,{\rm cl}}_{\mu\mu}$ does not. At finite temperatures of our interest, $\xi_s$ is finite, which means that in the whole system, there are domains of side length $\xi_s$ having different orientations of the 120$^\circ$ spin plane. Thus, at finite temperatures, the observable is the conductivity averaged over $\sigma^{s_{xz},{\rm cl}}_{\mu\mu}$, and $\sigma^{s_y,{\rm cl}}_{\mu\mu}$ and $\sigma^{s,{\rm cl}}_{\mu\mu} = \sigma^{s_{xz},{\rm cl}}_{\mu\mu} + \sigma^{s_y,{\rm cl}}_{\mu\mu} \sim const \,  T/\alpha_d + T \,\alpha_d \, \xi_s$ should correspond to the numerically obtained spin-current conductivity in the main text.
In contrast to the thermal conductivity $\kappa_{\mu\nu}^{\rm cl}$, the spin-current conductivity $\sigma^{s,{\rm cl}}_{\mu \mu}$ contains a term proportional to the spin correlation length $\xi_s \sim a \exp[b_H |J|/T]$. 
With decreasing temperature, the magnon damping $\alpha_d$ goes to zero, while $\xi_s$ increases, so that the temperature dependence of $\sigma^{s,{\rm cl}}_{\mu\mu} \sim const \, T/\alpha_d +  T \,\alpha_d \, \xi_s$ is not so trivial. Such a situation is in sharp contrast to the unfrustrated Heisenberg antiferromagnet on the square lattice in which the analytically obtained spin-current conductivity, $\sigma^{s, {\rm cl}}_{\mu\mu} \sim const \,  T/\alpha_d + T \xi_s/\alpha_d \sim T \xi_s/\alpha_d$, unambiguously increases in a monotonic manner toward $T=0$ \cite{TransportXXZ_AK_19}. The origin of the difference consists in the fact that the magnon velocities at the gapless points are the same in the unfrustrated square-lattice case, while not in the frustrated triangular-lattice case, i.e., $v_0 > v_{\pm{\bf Q}}$. Indeed, if $v_0 = v_{\pm{\bf Q}}$ were satisfied in the triangular-lattice system, the contribution proportional to $\xi_s$ in Eq. (\ref{eq:conductivity_classical_spin}) should involve not only a term of the form $T \, \alpha_d \xi_s$ but also a term of the form $T \, \xi_s/\alpha_d$, and the latter should be dominant in the low-temperature limit. 

So far, we have not discussed the specific temperature dependence of the magnon damping $\alpha_d$. In the case of the square-lattice antiferromagnets, the damping due to multi-magnon scatterings has been calculated in Refs. \cite{MagnonDamping_Tyc_89, MagnonDamping_Harris_71}, and the temperature dependence of $\alpha_d$ essentially follows the $T^2$ form, i.e., $\alpha_d \propto T^2$, which results from the leading-order scattering process involving {\it four} magnons. In the present triangular-lattice case, on the other hand, the leading-order scattering process involves {\it three} magnons, reflecting the non-collinearity of the 120$^\circ$ spin structure \cite{LSWT_rotation_Chubukov_jpcm_94, LSWT_rotation_Chernyshev_prb_09}. To our knowledge, no analytical result on the temperature dependence of $\alpha_d$ is available for the triangular-lattice classical antiferromagnet. In order to discuss the exact temperature dependence of $\sigma^{s,{\rm cl}}_{\mu\mu} \sim const \,  T/\alpha_d + T \,\alpha_d \, \xi_s$, detailed calculations would be required for $\alpha_d$, but this issue is beyond the scope of the present work.   

\section{ Time correlation functions of the spin and thermal currents}
\begin{figure}[t]
\includegraphics[scale=0.8]{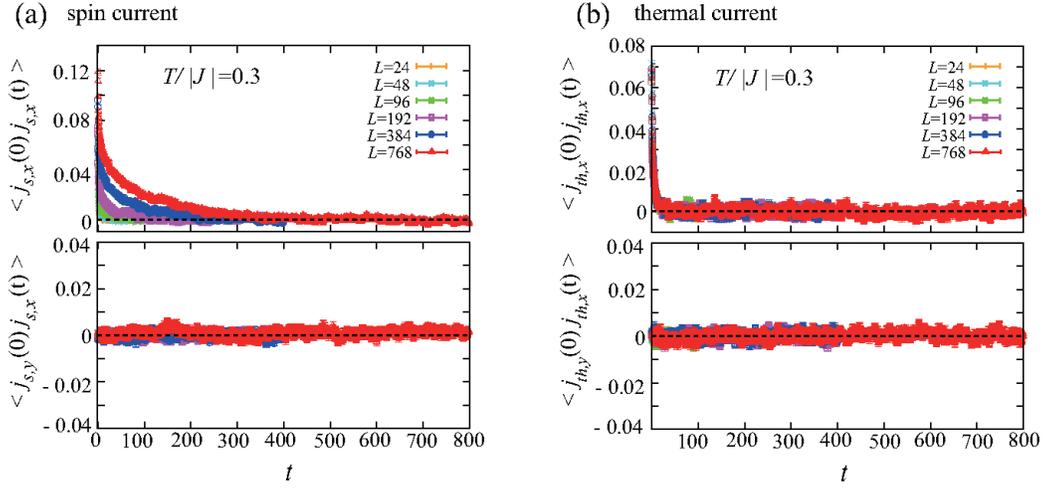}
\caption{The time correlation functions of the spin current $\langle j_{s,\mu}(0) j_{s,\nu}(t) \rangle$ (a) and the thermal current $\langle j_{th,\mu}(0) j_{th,\nu}(t) \rangle$ (b) at $T/|J|=0.3$, where the former and the latter are measured in units of $|J|^2$ and $|J|^4$, respectively. The upper and lower panels show the longitudinal ($\mu=\nu=x$) and transverse ($\mu=y, \, \nu=x$) correlations, respectively. The time $t$ is measured in units of $|J|^{-1}$. \label{fig:JsJth_tdep}}
\end{figure}
The longitudinal ($\mu=\nu=x$) and transverse ($\mu=y, \, \nu=x$) components of the time correlation function of the spin current $\langle j_{s,\mu}(0) j_{s,\nu}(t) \rangle$ and the thermal current $\langle j_{th,\mu}(0) j_{th,\nu}(t) \rangle$ at $T/|J|=0.3$ slightly above the $\mathbb{Z}_2$ vortex transition temperature $T_v/|J|=0.285$ are shown in Fig. \ref{fig:JsJth_tdep}. The upper panel of Fig. \ref{fig:JsJth_tdep} (a) corresponds to the longer-time view of the inset of Fig. 3 (a) in the main text. One can see that the transverse correlation is absent for both spin and thermal transports. The longitudinal time correlation of the spin current has a long-time relaxation showing a remarkable system-size dependence, while that of the thermal current has a relatively short-time relaxation independent of the system-size, suggesting that the spin and thermal currents are carried by different kinds of magnetic excitations.

\section{ Diffusive motion of the $\mathbb{Z}_2$-vortex}
\begin{figure}[t]
\includegraphics[scale=0.6]{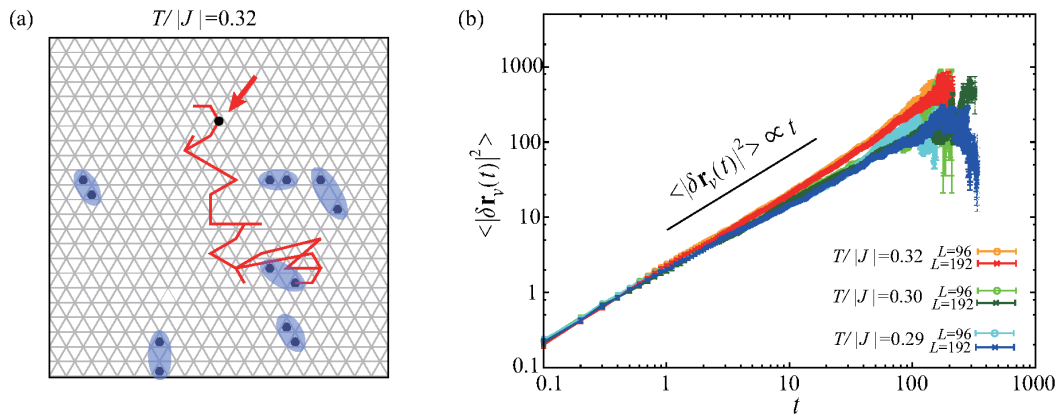}
\caption{(a) A snapshot of the $\mathbb{Z}_2$-vortex-core distribution taken in the MC simulation at $T/|J|=0.32$, and the trajectory of a free-vortex motion obtained in the spin-dynamics simulation, where black dots represent vortex cores and a red zigzag line represent the trajectory of a free-vortex core indicated by the red arrow. (b) Log-log plot of the time dependence of the mean square displacement $\langle |\delta {\bf r}_v(t)|^2 \rangle$ of the free $\mathbb{Z}_2$ vortex obtained at $T/|J|=0.32$ (reddish colored data), $T/|J|=0.3$ (greenish colored data), and $T/|J|=0.29$ (blueish colored data). The time $t$ is measured in units of $|J|^{-1}$. \label{fig:Z2_diffusive}}
\end{figure}

In order to characterize the motion of the free $\mathbb{Z}_2$ vortex, we shall calculate the mean square displacement $\langle |\delta {\bf r}_v(t)|^2 \rangle$ of the free-vortex core by tracing it in the spin-dynamics simulation. First, we shall pick up a target free $\mathbb{Z}_2$ vortex from the whole system. For an initial ($t=0$) equilibrium spin configuration taken in the MC simulation at a fixed temperature, one can obtain an associated vortex-core distribution by using the method addressed in Ref. \cite{Z2_Kawamura_84}. An example of the vortex-core distribution obtained in this way is shown in Fig. \ref{fig:Z2_diffusive} (a). Since a pair of the $\mathbb{Z}_2$ vortices whose core distance is shorter than 2$a$ ($a$ is the lattice constant) can be annihilated by a single spin flip, such a vortex pair is topologically unstable and can be regarded as a kind of spin-wave modes. In Fig. \ref{fig:Z2_diffusive} (a), such vortex pairs are represented by colored ellipses. By eliminating the vortex pairs, we obtain topologically stable free vortices. Then, as a target free $\mathbb{Z}_2$ vortex, we pick up the most isolated vortex which has the longest core distance to the NN free vortex. Suppose that the core position of the isolated vortex is denoted by ${\bf r}_v(0)$. Now, we consider the time evolution of the core position of the isolated vortex ${\bf r}_v(t)$. At the next time step ($t=\delta t$) in the spin-dynamics simulation, we calculate the vortex-core distribution and regard the position of the nearest vortex core to ${\bf r}_v(0)$ as the present position of the target vortex ${\bf r}_v(\delta t)$. Thus, the target vortex may stay there or move in a few lattice spacing. By performing this procedure at every time step, one obtains the trajectory of the target vortex ${\bf r}_v(t)$. If we cannot find any vortex core within a searching area of radius $R$ from the previous core position, we regard that the target vortex collides with an other free vortex and is pair-annihilated, and terminate the time evolution. The so-obtained trajectory of the isolated free $\mathbb{Z}_2$ vortex is represented by a zigzag red line in Fig. \ref{fig:Z2_diffusive} (a). One can see that the vortex motion is similar to Brownian motion, suggesting that it is diffusive. In order to quantitatively discuss the diffusive character of the vortex motion, we examine the time dependence of the mean square displacement $\langle |\delta {\bf r}_v(t)|^2 \rangle$ of the isolated free $\mathbb{Z}_2$ vortex, where $\delta {\bf r}_v(t)={\bf r}_v(t)-{\bf r}_v(0)$. Figure \ref{fig:Z2_diffusive} (b) shows $\langle |\delta {\bf r}_v(t)|^2 \rangle$ as a function of the time $t$ at various temperatures slightly above the $\mathbb{Z}_2$ vortex transition temperature $T_v/|J|=0.285$, where the radius of the searching area $R$ is set to be $R=5a$. As one can see from Fig. \ref{fig:Z2_diffusive} (b), $\langle |\delta {\bf r}_v(t)|^2 \rangle$ is roughly proportional to the time $t$, suggestive of the almost diffusive vortex motion. Although the exponent $x$ of the power-law form of $\langle |\delta {\bf r}_v(t)|^2 \rangle \propto t^x$ depends slightly on temperature, it also depends weakly on the radius of the searching area $R$. We note, however, that even for $R=2a$, the roughly diffusive character, i.e., $\langle |\delta {\bf r}_v(t)|^2 \rangle \propto t$, is confirmed to be unchanged.    

\section{The interface effect on the detection of the spin current}
As our focus in the present paper is on the bulk spin-current conductivity, the non-local measurement of the spin current is suitable for the investigation of this bulk transport property. In contrast, the measurement of the spin current locally generated at the interface with the use of the longitudinal spin Seebeck effect should be insensitive to the spin-current conductivity in the bulk away from the interface. Thus, we shall consider the non-local measurement.

In the main text, we have shown that the longitudinal spin-current conductivity in the bulk magnet exhibits a divergence at $T_v$. This means that in non-local measurement of the spin transport, a spin current injected by using, for example, the spin pumping effect, flows in a long distance in the bulk magnet near $T_v$, leading to a large spin accumulation at the opposite side of the sample. This spin accumulation could be measured by attaching a normal metal to the bulk magnet and detecting the inverse spin-Hall signal. In general, however, the signal strength depends on not only the amount of the spin accumulation but also the degree of its diffusion into the spin-Hall metal, i.e., the spin-mixing conductance (SMC) at the interface between the magnet and the metal. In the spin-current-injection sector, SMC is also important. Since as already mentioned, once the spin current is injected, the spin accumulation should become very large at $T_v$, the issue is the temperature dependence of SMC's at the interfaces. 

Following Refs. \cite{SMC_Takahashi_10, pmSSE_Adachi_19}, we assume that the spin diffusion at the interface occurs via the $s$-$d$ coupling of the form $J_{sd} \mbox{\boldmath$\sigma$}_i \cdot {\bf S}_i$, where ${\bf S}_i$ is a localized spin in the magnet and $\mbox{\boldmath$\sigma$}_i$ is a spin of a conduction electron in the normal metal. By using the perturbative expansion with respect to $J_{sd}$, one finds that the dynamical magnetic susceptibility $\chi^{+-}_m({\bf q},\omega)= i \int_0^\infty dt \, e^{i\omega t} \langle [S^-_{\bf q}(t), S^+_{-{\bf q}}(0)]\rangle $ is related to the spin current at the interface \cite{SMC_Takahashi_10}. Since the candidate magnets of our interest satisfy the fundamental property for the $\mathbb{Z}_2$-vortex transition, namely, any long-range magnetic order is absent down to the lowest temperature, the dynamical magnetic susceptibility $\chi^{+-}_m({\bf q},\omega)$ should not exhibit a critical behavior at any temperature. Indeed, in the candidate antiferromagnets NiGa$_2$S$_4$ \cite{NiGa2S4_Takeya_prb_08} and NaCrO$_2$ \cite{NaCrO2_Olariu_pbl_06}, a critical behavior characteristic of a second-order magnetic transition is {\it not} observed in the NMR experiment which is responsible for $\chi^{+-}_m({\bf q},\omega)$. Thus, in these compounds, the temperature dependence of the interface effect or SMC would be relatively weak within the $s$-$d$ coupling model, although its details may depend also on the electronic structure in the metal side. This suggests that if one detects a large inverse spin-Hall signal at a putative $\mathbb{Z}_2$-vortex transition temperature, it could be attributed to the divergent enhancement of the bulk spin-current conductivity $\sigma^s_{\mu\mu}$ and the resultant spin accumulation rather than the interface effect.

Here, we will briefly comment on the measurement of the spin current with the use of the longitudinal spin Seebeck effect. Since the $\mathbb{Z}_2$-vortex transition occurs in the magnetically disordered phase, an experimental technique available in this temperature regime would be the paramagnetic spin Seebeck effect \cite{pmSSE_Adachi_19, pmSSE_Wu_15, Spincurrent-mag_Li_19} in which an external magnetic field needs to be applied. In this case, it is theoretically shown that the intensity of the spin current at the interface is proportional to the induced magnetic moment, i.e., the magnetic susceptibility $\chi_m$ and the field strength \cite{pmSSE_Adachi_19}. In a standard collinear antiferromagnet, $\chi_m$ shows a cusp at the magnetic transition, whereas in the present frustrated antiferromagnet, $\chi_m$ only shows a weak essential singularity at $T_v$ \cite{Z2_Kawamura_84, Z2_Kawamura_10}. Thus, in the longitudinal spin Seebeck experiment, the signature of the phase transition can be captured in the collinear antiferromagnet, but not in the present frustrated system, which calls for the non-local measurement of the spin current for the detection of the $\mathbb{Z}_2$ vortex transition.


\begin{thebibliography}{100}
\bibitem{Z2_Kawamura_84} H. Kawamura and S. Miyashita, J. Phys. Soc. Jpn. {\bf 53}, 4138 (1984).
\bibitem{Z2_Southern_95} B. W. Southern and H-J. Xu, Phys. Rev. B {\bf 52}, 3836(R) (1995).
\bibitem{Z2_Wintel_95} M. Wintel, H. U. Everts, and W. Apel, Phys. Rev. B {\bf 52}, 13480 (1995).
\bibitem{Z2_Kawamura_10} H. Kawamura, A. Yamamoto, and T. Okubo, J. Phys. Soc. Jpn. {\bf 79}, 023701 (2010).
\bibitem{Z2_Kawamura_11} H. Kawamura, J. Phys. Conf. Ser. {\bf 320}, 012002 (2011).
\bibitem{Sqomega_Okubo_jpsj_10} T. Okubo and H. Kawamura, J. Soc. Phys. Jpn. {\bf 79}, 084706 (2010).

\bibitem{NiGa2S4_Nakatsuji_science_05} S. Nakatsuji, Y. Nambu, H. Tonomura, O. Sakai, S. Jonas, C. Broholm, H. Tsunetsugu, Y. Qiu, and Y. Maeno, Science {\bf 309}, 1697 (2005).
\bibitem{NiGa2S4_Nambu_jpsj_06} Y. Nambu, S. Nakatsuji, and Y. Maeno, J. Phys. Soc. Jpn. {\bf 75}, 043711 (2006).
\bibitem{NiGa2S4_Nambu_prl_08} Y. Nambu, S. Nakatsuji, Y. Maeno, E. K. Okudzeto, and J. Y. Chan, Phys. Rev. Lett. {\bf 101}, 207204 (2008).
\bibitem{NiGa2S4_Takeya_prb_08} H. Takeya, K. Ishida, K. Kitagawa, Y. Ihara, K. Onuma, Y. Maeno, Y. Nambu, S. Nakatsuji, D. E. MacLaughlin, A. Koda, and R. Kadono, Phys. Rev. B {\bf 77}, 054429 (2008).
\bibitem{NiGa2S4_MacLaughlin_prb_08} D. E. MacLaughlin, Y. Nambu, S. Nakatsuji, R. H. Heffner, L. Shu, O. O. Bernal, and K. Ishida, Phys. Rev. B {\bf 78}, 220403(R) (2008).
\bibitem{NiGa2S4_Yaouanc_prb_08} A. Yaouanc, P. Dalmas de Reotier, Y. Chapuis, C. Marin, G. Lapertot, A. Cervellino, and A. Amato, Phys. Rev. B {\bf 77}, 092403 (2008).
\bibitem{NiGa2S4_Yamaguchi_prb_08} H. Yamaguchi, S. Kimura, M. Hagiwara, Y. Nambu, S. Nakatsuji, Y. Maeno, and K. Kindo, Phys. Rev. B {\bf 78}, 180404(R) (2008).
\bibitem{NiGa2S4_Yamaguchi_jpsj_10} H. Yamaguchi, S. Kimura, M. Hagiwara, Y. Nambu, S. Nakatsuji, Y. Maeno, A. Matsuo, and K. Kindo, J. Phys. Soc. Jpn. {\bf 79}, 054710 (2010).
\bibitem{NiGa2S4_Nakatsuji_review_10} S. Nakatsuji, Y. Nambu, and S. Onoda, J. Phys. Soc. Jpn. {\bf 79}, 011003 (2010).
\bibitem{NiGa2S4_Nambu_review_10} Y. Nambu and S. Nakatsuji, J. Phys.: Condens. Matter {\bf 23}, 164202 (2011).
\bibitem{NiGa2S4_Stock_prl_10} C. Stock, S. Jonas, C. Broholm, S. Nakatsuji, Y. Nambu, K. Onuma, Y. Maeno, and J.-H. Chung, Phys. Rev. Lett. {\bf 105}, 037402 (2010).
\bibitem{NiGa2S4_Nambu_prl_15} Y. Nambu, J. S. Gardner, D. E. MacLaughlin, C. Stock, H. Endo, S. Jonas, T. J. Sato, S. Nakatsuji, and C. Broholm, Phys. Rev. Lett. {\bf 115}, 127202 (2015).

\bibitem{FeGa2S4_Zhao_prb_12} Songrui Zhao, P. Dalmas de Reotier, A. Yaouanc, D. E. MacLaughlin, J. M. Mackie, O. O. Bernal, Y. Nambu, T. Higo, and S. Nakatsuji, Phys. Rev. B {\bf 86}, 064435 (2012).
\bibitem{FeGa2S4_Reotier_prb_12} P. Dalmas de Reotier, A. Yaouanc, D. E. MacLaughlin, Songrui Zhao, T. Higo, S. Nakatsuji, Y. Nambu, C. Marin, G. Lapertot, A. Amato, and C. Baines, Phys. Rev. B {\bf 85}, 140407(R) (2012).

\bibitem{NaCrO2_Olariu_pbl_06} A. Olariu, P. Mendels, F. Bert, B. G. Ueland, P. Schiffer, R. F. Berger, and R. J. Cava, Phys. Rev. Lett. {\bf 97}, 167203 (2006).
\bibitem{NaCrO2_Hsieh_physicaB_08} D. Hsieh, D. Qian, R. F. Berger, R. J. Cava, J. W. Lynn, Q. Huang, and M. Z. Hasan, Physica B {\bf 403}, 1341 (2008).
\bibitem{NaCrO2_Hsieh_jpcs_08} D. Hsieh, D. Qian, R. F. Berger, R. J. Cava, J. W. Lynn, Q. Huang, and M. Z. Hasan, J. Phys. Chem. Solids {\bf 69}, 3174 (2008).
\bibitem{NaCrO2_Hemmida_prb_09} M. Hemmida, H. A. Krug von Nidda, N. Buttgen, A. Loidl, L. K. Alexander, R. Nath, A. V. Mahajan, R. F. Berger, R. J. Cava, Yogesh Singh, and D. C. Johnston, Phys. Rev. B {\bf 80}, 054406 (2009).

\bibitem{KCrO2_Soubeyroux_79} J. L. Soubeyroux, D. Fruchart, C. Delmas, and G. Le. Flem, J. Mag. Mag. Mater. {\bf 14}, 159 (1979).
\bibitem{KCrO2_Xiao_prb_13} F. Xiao, T. Lancaster, P. J. Baker, F. L. Pratt, S. J. Blundell, J. S. Moller, N. Z. Ali, and M. Jansen, Phys. Rev. B {\bf 88}, 180401(R) (2013). 

\bibitem{AAg2CrV2O8_Tapp_prb_17} J. Tapp, C. R. Dela Cruz, M. Bratsch, N. E. Amuneke, L. Postulka, B. Wolf, M. Lang, H. O. Jeschke, R. Valenti, P. Lemmens, and A. Moller, Phys. Rev. B {\bf 96}, 064404 (2017). 

\bibitem{Spincurrent-mag_Frangou_16} L. Frangou, S. Oyarzun, S. Auffret, L. Vila, S. Gambarelli, and V. Baltz, Phys. Rev. Lett. {\bf 116}, 077203 (2016).
\bibitem{Spincurrent-mag_Qiu_16} Z. Qiu, J. Li, D. Hou, E. Arenholz, A. T. N'Diaye, A. Tan, K. Uchida, K. Sato,
S. Okamoto, Y. Tserkovnyak, Z. Q. Qiu, and E. Saitoh, nat. commun. {\bf 7}, 12670 (2016).
\bibitem{Spincurrent-mag_Wang_17} H. Wang, D. Hou, Z. Qiu, T. Kikkawa, E. Saitoh, and X. Jin, J. Appl. Phys. {\bf 122}, 083907 (2017).
\bibitem{Spincurrent-mag_Frangou_17} L. Frangou, G. Forestier, S. Auffret, S. Gambarelli, and V. Baltz, Phys. Rev. B {\bf 95}, 054416 (2017).
\bibitem{Spincurrent-mag_Gladii_18} O. Gladii, L. Frangou, G. Forestier, R. L. Seeger, S. Auffret, I. Joumard, M. Rubio-Roy, S. Gambarelli, and V. Baltz, Phys. Rev. B {\bf 98}, 094422 (2018).
\bibitem{Spincurrent-mag_Ou_18} Y. Ou, D. C. Ralph, and R. A. Buhrman, Phys. Rev. Lett. {\bf 120}, 097203 (2018).
\bibitem{Spincurrent-mag_Li_19} J. Li, Z. shi, V. H. Ortiz, M. Aldosary, C. Chen, V. Aji, P. Wei, and J. Shi, Phys. Rev. Lett. {\bf 122}, 217204 (2019).

\bibitem{SpinDyn_Huber_74} N. A. Lurie, D. L. Huber, and M. Blume, Phys. Rev. B {\bf 9}, 2171 (1974).  
\bibitem{SpinDyn_Jencic_prb_15} B. Jencic and P. Prelovsek, Phys. Rev. B {\bf 92}, 134305 (2015).  
\bibitem{MHall_Mook_prb_16} A. Mook, J. Henk, and I. Mertig, Phys. Rev. B {\bf 94}, 174444 (2016).   
\bibitem{MHall_Mook_prb_17} A. Mook, B. Gobel, J. Henk, and I. Mertig, Phys. Rev. B {\bf 95}, 020401(R) (2017).  
\bibitem{Thermal_Huber_ptp_68} D. L. Huber, Prog. Theor. Phys. {\bf 39}, 1170 (1968). 
\bibitem{SpinDyn_Zotos_prb_05} A. V. Savin, G. P. Tsironis, and X. Zotos, Phys. Rev. B {\bf 72}, 140402(R) (2005).  
\bibitem{SpinDyn_Kawasaki_67} K. Kawasaki, J. Phys. Chem. Solids, {\bf 28}, 1277 (1967).  
\bibitem{SpinDyn_Sentef_07} M. Sentef, M. Kollar, and A. P. Kampf, Phys. Rev. B {\bf 75}, 214403 (2007).  
\bibitem{SpinDyn_Pires_09} A. S. T. Pires and L. S. Lima, Phys. Rev. B {\bf 79}, 064401 (2009)  
\bibitem{SpinDyn_Chen_13} Z. Chen, T. Datta, and D. Yao, Eur. Phys. J. B {\bf 86}, 63 (2013). 

\bibitem{TransportXXZ_AK_19} K. Aoyama and H. Kawamura, Phys. Rev. B {\bf 100}, 144416 (2019).

\bibitem{KuboFormular_Kubo_57} R. Kubo, J. Phys. Soc. Jpn. {\bf 12}, 570 (1957).

\bibitem{Symplectic_Krech_98} M. Krech, A. Bunker, and D.P. Landau, Comput. Phys. Commun. {\bf 111}, 1-13 (1998).
\bibitem{Symplectic_Furuya_11} S. C. Furuya, M. Oshikawa, and I. Affleck, Phys. Rev. B {\bf 83}, 224417 (2011).

\bibitem{Suppl} K. Aoyama and H. Kawamura, Supplemental Material which includes Refs. \cite{Heisenberg_Polyakov_75,MagnonDamping_Tyc_89,LSWT_rotation_Chubukov_jpcm_94,LSWT_rotation_Chernyshev_prb_09,book_AGD,MagnonGreen_Yamaguchi_17,MagnonTrans_Tatara_15,MagnonDamping_Harris_71}.
\bibitem{Heisenberg_Polyakov_75} A. M. Polyakov, Phys. Lett. B {\bf 59}, 79 (1975).
\bibitem{MagnonDamping_Tyc_89} S. Tyc and B. I. Halperin, Phys. Rev. B {\bf 42}, 2096 (1990).
\bibitem{LSWT_rotation_Chubukov_jpcm_94} A. V. Chubukov, S. Sachdev, and T. Senthil, J. Phys.: Condens. Matter {\bf 6}, 8891 (1994). 
\bibitem{LSWT_rotation_Chernyshev_prb_09}  A. L. Chernyshev, M. E. Zhitomirsky, Phys. Rev. B {\bf 79}, 144416 (2009).
\bibitem{book_AGD} A. A. Abrikosov, L. P. Gorkov, and I. E. Dzyaloshinski, {\it Methods of Quantum Field Theory in Statistical Physics}, (Dover Publications, New York, 1963). 
\bibitem{MagnonGreen_Yamaguchi_17} T. Yamaguchi and H. Kohno, J. Phys. Soc. Jpn. {\bf 86}, 063706 (2017).
\bibitem{MagnonTrans_Tatara_15} G. Tatara, Phys. Rev. B {\bf 92}, 064405 (2015).
\bibitem{MagnonDamping_Harris_71} A. B. Harris, D. Kumar, B. I. Halperin, and P. C. Hohenberg, Phys. Rev. B {\bf 3}, 961 (1971).

\bibitem{diffusive_suppl} See Supplemental Material for the numerical result on the diffusive motion of the $\mathbb{Z}_2$ vortex.
\bibitem{nonlocal_Kajiwara_10} Y. Kajiwara, K. Harii, S. Takahashi, J. Ohe, K. Uchida, M. Mizuguchi, H. Umezawa, H. Kawai, K. Ando, K. Takanashi, S. Maekawa, and E. Saitoh, Nature {\bf 464}, 262 (2010). 
\bibitem{nonlocal_Giles_15} B. L. Giles, Z. Yang, J. S. Jamison, and R. C. Myers, Phys. Rev. B {\bf 92}, 224415 (2015).
\bibitem{nonlocal_Cornelissen_15} L. J. Cornelissen, J. Liu, R. A. Duine, J. Ben Youssef, and B. J. van Wees, Nat. Phys. {\bf 11}, 1022 (2015).

\bibitem{interface_suppl} See Supplemental Material for the effect of the interface with the spin-Hall metal, which includes Refs. \cite{SMC_Takahashi_10,pmSSE_Adachi_19,pmSSE_Wu_15}.
\bibitem{SMC_Takahashi_10} S. Takahashi, E. Saitoh, and S. Maekawa, J. Phys.: Conf. Ser. {\bf 200}, 062030 (2010).
\bibitem{pmSSE_Adachi_19} Y. Yamamoto, M. Ichioka, and H. Adachi, Phys. Rev. B {\bf 100}, 064419 (2019).
\bibitem{pmSSE_Wu_15} S. M. Wu, J. E. Pearson, and A. Bhattacharya, Phys.Rev. Lett. {\bf 114}, 186602 (2015).

\bibitem{SpinDyn_Bennett_65} H. S. Bennett and P. C. Martin, Phys. Rev. {\bf 138}, A 608 (1965).
\bibitem{SpinDyn_Kawasaki_68} K. Kawasaki, Prog. Theor. Phys. {\bf 39}, 1133 (1968).
\bibitem{SpinDyn_Hohenberg_77} P. C. Hohenberg and B. I. Halperin, Pev. Mod. Phys. {\bf 49}, 435 (1977).

\end{thebibliography}
\end{document}